\documentclass[pdflatex,sn-mathphys-num]{sn-jnl}

\usepackage{amsmath,amssymb,amsfonts,amsthm}
\usepackage{mathrsfs}
\usepackage[title]{appendix}
\usepackage{float}
\usepackage[font=small,labelfont=bf]{caption}
\usepackage{comment}
\usepackage{ulem}

\usepackage{graphicx}
\usepackage[table]{xcolor}

\usepackage{booktabs}
\usepackage{multirow}
\usepackage{tabularx}
\newcolumntype{L}[1]{>{\hsize=#1\hsize\raggedright\arraybackslash}X}%
\newcolumntype{R}[1]{>{\hsize=#1\hsize\raggedleft\arraybackslash}X}%
\newcolumntype{C}[1]{>{\hsize=#1\hsize\centering\arraybackslash}X}

\usepackage[separate-uncertainty=true, %
            multi-part-units=single, %
            range-units=single]{siunitx}

\DeclareMathOperator*{\argmax}{argmax}
\newcommand{\braket}[1]{\langle #1 \rangle}
\DeclareSIUnit{\flop}{flop}
\DeclareSIUnit{\node}{node}
\DeclareSIUnit{\neuron}{neuron}
\DeclareSIUnit{\ATP}{ATP}
\DeclareSIUnit{\ATPW}{\text{ATP}\text{-}\text{W}}
\DeclareSIUnit{\cal}{cal}
\DeclareSIUnit{\spike}{spike}
\DeclareSIUnit{\corehour}{corehour}

\newcommand{\ie}{\textit{i.e.}}
\newcommand{\eg}{\textit{e.g.}}

\definecolor{bluePier}{rgb}{0.0, 0.0, 1.0}

\definecolor{purpleElena}{rgb}{1.0, 0.0, 1.0}

\definecolor{applegreen}{rgb}{0.55, 0.71, 0.0}

\definecolor{brownLeo}{rgb}{0.59, 0.33, 0.2}

\begin{document}


\title[Full Plasticity]{Unsupervised sleep-like intra- and inter-layer plasticity categorizes and improves energy efficiency in a multilayer spiking network}


\author[1,a]{\fnm{Leonardo} \sur{Tonielli}}

\author[1,a,b]{\fnm{Cosimo} \sur{Lupo}}

\author[1]{\fnm{Elena} \sur{Pastorelli}}

\author[1]{\fnm{Giulia} \sur{De Bonis}}

\author[1]{\fnm{Francesco} \sur{Simula}}

\author[1]{\fnm{Alessandro} \sur{Lonardo}}

\author[1]{\fnm{Pier Stanislao} \sur{Paolucci}}

\affil[1]{\orgname{Istituto Nazionale di Fisica Nucleare}, \orgdiv{Sezione di Roma}, \orgaddress{\street{Piazzale Aldo Moro 2}, \city{Roma}, \postcode{00185}, \country{Italy}}}

\affil[a]{These authors contributed equally}
\affil[b]{Corresponding author: \href{mailto:cosimo.lupo89@gmail.com}{cosimo.lupo89@gmail.com}}


\abstract{
    Sleep is thought to support memory consolidation and the recovery of optimal energetic regime by reorganizing synaptic connectivity, yet how plasticity across hierarchical brain circuits contributes to abstraction and energy efficiency remains unclear.
    Here we study a spiking multi-layer network alternating wake-like and deep-sleep-like states, with state-dependent dendritic integration and synaptic plasticity in a biologically inspired thalamo-cortical framework.
    During wakefulness, the model learns from few perceived examples, while during deep sleep it undergoes spontaneous replay driven by slow oscillations. Plasticity enabled not only within intra-layer connections, but also in inter-layer pathways, is critical for memory consolidation and energetic downshift.
    Compared to restricted plasticity, full inter-layer plasticity yields higher post-sleep visual classification accuracy and promotes the emergence of sharper class-specific associations.
    Furthermore, we introduce a biophysically grounded estimator of metabolic power expressing network energy consumption in ATP units, partitioned into baseline, synaptic maintenance, action potential, and transmission costs.
    We find that inter-layer plasticity in sleep leads to a larger reduction in firing rates, synaptic strength and synaptic activity, corresponding to a substantially larger decrease in power consumption.
    This work suggests promising elements to be integrated in neuromorphic/energy-efficient AI learning systems, supported by brain state-specific apical mechanisms.
}


\maketitle

\section{Introduction}
\label{sec:intro}

\begin{figure}[!b]
    \centering
    \includegraphics[width=\textwidth]{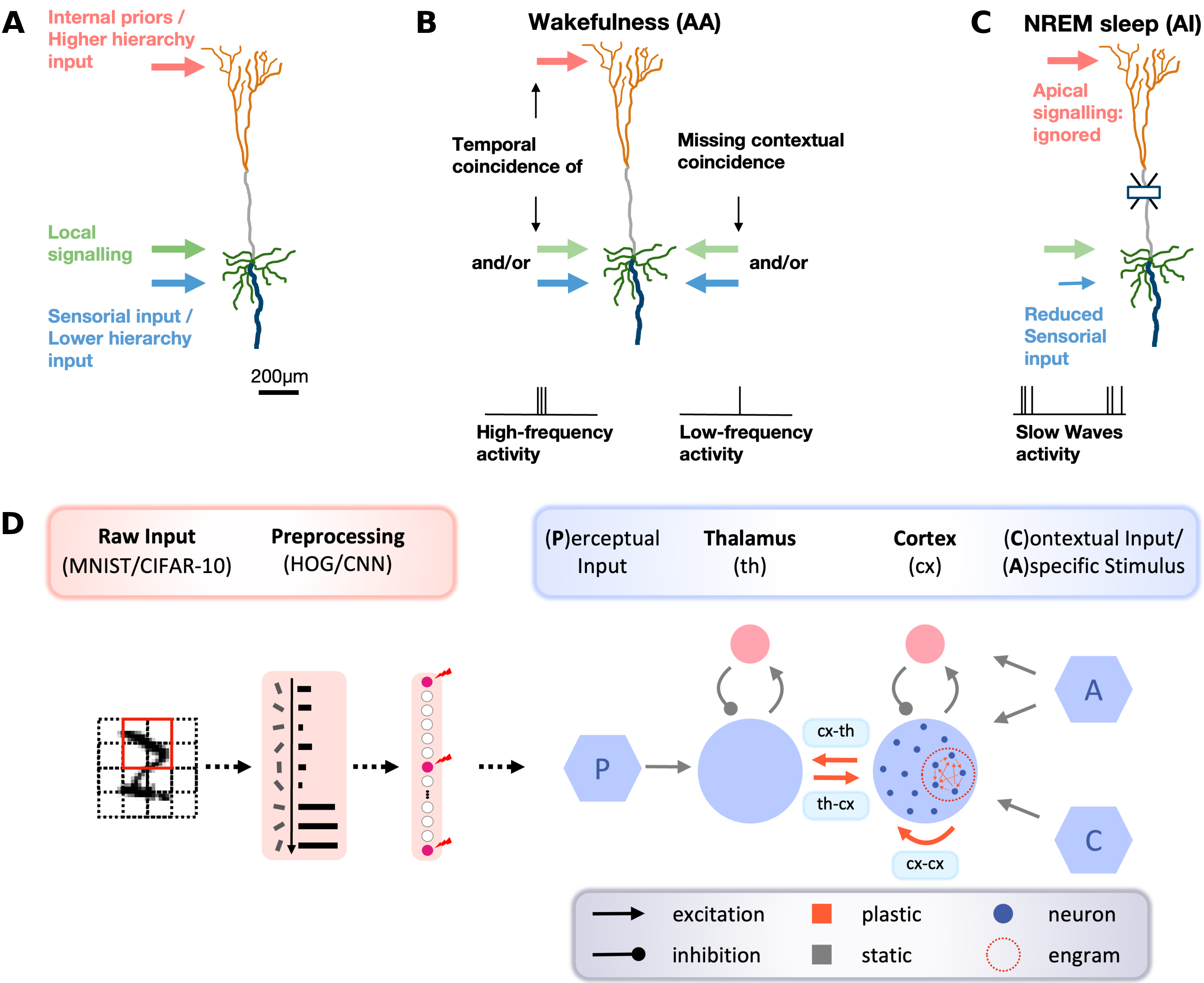}
    \caption{
        \textbf{A biologically grounded multi-layer model with inter- and intra-layer recurrence for sleep and learning}.
        \textbf{A)} Layer-5 cortical pyramidal cell integrating contextual signals (red) with perceptual (blue) and/or local (green) signals.
        \textbf{B)} Apical amplification: during wakefulness, this mechanism enables the amplification of the spiking activity if apical signal is received in temporal coincidence with (peri-)somatic inputs.
        \textbf{C}) Apical isolation: during deep sleep, this mechanism prevents from apical signals to reach the soma.
        Panels A-C adapted from~\citet{pastorelli2025simplified}.
        \textbf{D)} Two-layer excitatory-inhibitory spiking neural network sustaining learning and sleep through apical mechanisms. Circles: neuronal populations, excitatory and inhibitory; hexagons: inputs to the network, encoded by the rates of Poisson processes. Either HOG-filtered MNIST digits, or CNN-processed CIFAR-10 images, are injected into the thalamus as perceptual (P) signals. Inter-layer (th$\to$cx and cx$\to$th) and recurrent (cx$\to$cx) excitatory connections are plastic (red), while others remain static (grey). Contextual (C) stimuli, administered during training, are necessary for engrams creation, while aspecific (A) random signals are necessary for inducing NREM-like slow oscillations during sleep.
    }
    \label{fig:net_architecture}
\end{figure}

Sleep is known to be essential for optimizing cognitive functions, reducing post-sleep local energy consumption and balancing energetic needs across brain areas \cite{diekelmann2010memory, killgore2010effects}, although a comprehensive understanding of its underlying mechanisms is still lacking.
Notably, the human brain consumes \qty{20}{\percent} of the resting metabolism, although it accounts for only \qty{2}{\percent} of the body weight, using up to half of the available nutrients during early childhood \cite{padamsey2023paying, jamadar2025metabolic}. The modelling of the energetic effects of sleep-induced reorganization of its connections is thus relevant, though yet to be addressed.
In the last three decades, mounting experimental evidence \cite{larkum1999new, larkum2009synaptic, larkum2013cellular} has fostered the conceptual formalization of cellular and architectural principles believed to shape both micro- and macro-scale neural activity during waking and sleep states.
At the cellular level, such principles leverage the elongated structure of pyramidal neurons, specifically in layer~\num{5} of the cortex (Fig.~\ref{fig:net_architecture}A). Such neurons exhibit clearly distinct apical dendritic tuft and (peri-)somatic dendritic zone, enabling a rich compartment-specific and integrated dynamics, and allowing for nonlinear computation directly at the neuron level \cite{poirazi2020illuminating, chavlis2025dendrites}. At the macro-scale, inter-areal connections target the apical tuft, supporting the identification of signals carrying contextual information and the inner status of other brain areas, while the perceptual stimulus mainly target the (peri-)somatic zone.
Grounded in this structure, two key mechanisms come into play, responsible for expressing learning capabilities of biological neuronal networks. \textit{Apical amplification}~\cite{phillips2016effects, phillips2023cooperative, phillips2025cellular} is believed to play a crucial role during wakefulness~\cite{aru2020cellular}, inducing high-frequency bursting in experience-dependent minorities of neurons, thereby enabling the detection of temporal coincidences between internal priors -- mainly targeting the apical tuft -- and perceptual signals -- impinging on the (peri-)somatic zone (Fig.~\ref{fig:net_architecture}B). During learning, this mechanism supports the formation of novel experience-specific neuronal assemblies of small size, while avoiding catastrophic forgetting of existing ones.
In contrast, during deep sleep, \textit{apical isolation} is thought to decouple somatic and apical compartments~\cite{suzuki2020general}. Since inter-areal contextual information is conveyed either by projections from non-specific thalamic nuclei or by direct cortico-cortical connections both targeting the apical tuft, apical isolation restricts processing to local interactions, causing neurons to ignore inter-areal signals (Fig.~\ref{fig:net_architecture}C). Apical isolation should thus facilitate local synaptic optimization within cortical areas, with cognitive and energetic benefits observed during post-sleep wakefulness.

The corpus of experimental evidence has prompted the modelling -- via plastic spiking network simulations -- of state-dependent brain mechanisms that enable the beneficial cognitive and energetic effects of sleep through the reorganization of engrams created during wakefulness~\cite{josselyn2015finding}.
A first example is presented in~\citet{wei2016synaptic}, where a thalamo-cortical system is implemented to model a single cortical column and its core and matrix thalamic inputs, using Hodgkin-Huxley point neurons and spike-timing-dependent plasticity (STDP). The network consists of chains of neurons with short-range local connectivity and both feed-forward and backward inter-layer connections. The goal is to measure the plasticity effects of deep-sleep-like endogenous rhythms on both cortical activity and synaptic structure. Such a model exhibits the capability to reproduce homogeneous slow oscillations (SO) of the entire network; in addition, the simulated dynamics promotes spontaneous replay of specific cortical neural sequences and the emergence of neural ensembles which initiate the up states. Synaptic changes are also reported, involving both long-term potentiation (LTP) and long-term depression (LTD) of intra-layer cortico-cortical connections. However, the highly schematic representation of network geometry and the lack of focus on sleep-related effects on classification performances leave ample room for biologically grounded enhancements.

Another solution is illustrated in~\citet{golden2022sleep}, consisting of a three-layer feed-forward point-neuron spiking network that solves a foraging task through reinforcement learning, reducing it to a basic pattern recognition problem limited to only two patterns, composed of two pixels each. In this model, a REM-sleep-like phase is introduced as an offline stage during which the network is induced to replay previously learned feed-forward synaptic patterns, thereby mitigating the catastrophic forgetting that would otherwise occur during subsequent training on a new task. Such REM phase is induced by Poisson spike trains, so that the average firing rate approximates the one observed during the training stage. This model relies exclusively on feed-forward connections and employs STDP to drive synaptic changes during training and sleep phases. The learning rule explicitly integrates \textit{ad~hoc} regularization mechanisms to enforce synaptic and firing-rate homeostasis, as well as synaptic rescaling, without recourse to more biologically plausible mechanisms (as, \eg, spike-frequency adaptation and winner-take-all).

A more biologically grounded approach has been introduced in \citet{capone2019sleeplike}, that proposes a minimal thalamo-cortical plastic spiking model (ThaCo) capable of encoding, retrieving, and classifying visual stimuli. This framework employs a contextual signal to select example-specific cortical assemblies and a sleep-like regime that generates slow oscillations and endogenous replay. Through STDP-driven reactivation, the model exhibits the characteristic dual synaptic signature of biological sleep -- homeostatic LTD of example-specific cortico-cortical connections and associative LTP among cortical neurons representing the same category. These dynamics improve the network sensitivity to category-specific inputs and enhance task-related performance after sleep.
The ThaCo model has been extended in~\citet{golosio2021thalamocortical} to support fast incremental learning along multiple cycles of learning and sleep, under noisy perceptual and contextual conditions, while reproducing cortical firing-rate statistics consistent with physiology. In this work, non-rapid-eye-movement (NREM)-like dynamics further contributes to firing-rate homeostasis and reduced vulnerability to noise. However, in both studies, plasticity during sleep has been intentionally restricted to cortico-cortical synapses, leaving the role of thalamo-cortical and cortico-thalamic plasticity during sleep unexplored.

In the present study, we \textit{i)} further extend the two aforementioned thalamo-cortical models \cite{capone2019sleeplike, golosio2021thalamocortical} and \textit{ii)} introduce a novel methodology for the estimation of network energetic costs expressed in metabolic units, grounded on experimental findings available at cellular and areal scale.

The extension of the model is focused on the effects of sleep-induced multi-layer synaptic reorganization (in full-plasticity mode) on a classification task that is performed after learning an extremely low number of examples. Specifically, only three examples per class are used to train the network, drawn from two benchmark datasets: the \num{28}$\times$\num{28}-pixel handwritten digits from MNIST~\cite{deng2012mnist} and the \num{32}$\times$\num{32}-pixel RGB coloured images from CIFAR-10~\cite{krizhevsky2009cifar}.
We show that if plasticity is enabled across feed-forward, backward and recurrent connections, sleep optimizes thalamo-cortical perceptual representations in an entirely unsupervised manner. This naturally leads to the global homeostasis of the synaptic structure, compensated by the association of similar concepts, via LTD and LTP of thalamo-cortical synapses. Overall, this mechanism supports the normalization of internal representations formed during wakefulness, ultimately promoting the consolidation of learned memories. As a result, the network exhibits further enhancements during post-sleep wakefulness, characterized by both improved classification efficiency and reduced energetic costs.

In what follows, details are given on the efforts made in this work to introduce substantial improvements with respect to the above mentioned literature.
Compared to~\citet{wei2016synaptic}, the present model features homogeneous all-to-all connectivity for all types of connections. However, the apical amplification mechanism enables the emergence of distinct thalamo-cortical assemblies during the training phase. Importantly, our model simulates slow oscillations within neural ensembles that collectively give rise to delta-band oscillatory dynamics at the network level; this differs from \citet{wei2016synaptic}, where the primary purpose is to simulate spatially propagating waves across the network.

Differently from~\citet{golden2022sleep}, our model does not require additional \textit{ad hoc} regularization terms to achieve the benefits observed during both learning and sleep phases, since those benefits are endogenously generated through a pairwise non-linear time-additive Hebbian (NLTAH) STDP rule~\cite{guetig2003learning}. Furthermore,  plasticity affects the whole excitatory thalamo-cortical network. In addition, SO dynamics drives a mutual (co-)activation of cell assemblies, regulated by biologically plausible mechanisms like spike-frequency adaptation and winner-take-all. Overall, this results in the endogenous orchestration of neural ensembles, leading to spontaneously emerging cognitive and energetic benefits upon awakening.

In comparison with the first version of the ThaCo model \cite{capone2019sleeplike}, the present study shows that enabling thalamo-cortical and cortico-thalamic plasticity during sleep, in addition to the cortico-cortical one, leads to remarkably enhanced cognitive benefits. In particular, it significantly improves classification performance and promotes a global endogenous down-regulation of cortical activity levels during wakefulness.

Improving on prior metrics for the estimation of metabolic costs our study is, according to our knowledge, the first to specifically focus on the effect of plasticity in spiking networks. Starting from elementary metabolic processes, we introduce a practical ATP-calibrated estimator, thereby enabling the quantification of the energetic benefits of synaptic reorganization during endogenous, sleep-like dynamics.

Also, this work demonstrates a reorganization of both intra-layer (cortico-cortical) and inter-layer (cortico-thalamo-cortical) synapses in a hierarchical network set -- thanks to the emulation of apical isolation -- into spontaneous deep-sleep-like dynamics, after learning memories captured thanks to the emulation of apical amplification.

In summary, we show that enabling both intra- and inter-layer plasticity during a deep-sleep-like state, beneficially reorganizes cortical engrams, improving classification after few-shot learning and reducing the post-sleep metabolic cost of the network during the classification task.
The mechanism we investigate appears to be generalizable to any neighbouring pair of network layers, and supports both intra-layer recurrence, and unsupervised inter-layer feedback. However, in continuity with the approach we introduced with the ThaCo models, we decided in this work to forge it on representing a small cortical patch and its thalamic nuclei.

\section{Results}
\label{sec:results}


The novel model preserves the two canonical sleep signatures -- namely, reduction of the rate of occurrence of SO and firing-rate down-regulation, that align with \textit{in vivo} observations -- and extends prior models by admitting full cortico-cortical (cx$\to$cx), thalamo-cortical (th$\to$cx) and cortico-thalamic (cx$\to$th) plasticity during sleep, yielding measurable improvements in post-sleep classification accuracy and energy consumption.
Findings of the novel model are: \textit{i)} the cortical signatures of sleep are preserved and quantifiable, \textit{ii)} small but systematic inter-layer synaptic adjustments emerge, \textit{iii)} these modifications are associated with larger post-sleep gains in classification accuracy after few-shot training, and \textit{iv)} a stronger down-shift of post-sleep awake firing rates, together with a pronounced synaptic optimization, resulting into a more relevant reduction of the metabolic power, quantified via a novel biologically-grounded estimator.
Ablation comparisons (full plasticity vs. cx$\to$cx only) eventually point at cx$\to$th plasticity as a critical pathway for consolidating class-level synaptic structure, without compromising the homeostatic down-selection responsible for power optimization.


\subsection{Model setup}
\label{sec:results:setup}
Entering into the details -- fully unrolled in Sec.~\nameref{sec:methods} -- the model we implemented for studying the impact of sleep on cognitive and energetic performance is a two-layer plastic spiking network (Fig.~\ref{fig:net_architecture}D), in which all excitatory synapses -- both intra- and inter-layer -- are kept plastic during learning and sleep stages; conversely, to isolate the effects of sleep-induced synaptic changes on performance, synaptic plasticity is disabled during classification.

Both layers in the network are made of excitatory and inhibitory populations of conductance-based adaptive exponential leaky-integrate-and-fire point neurons. Inter-layer th$\to$cx and cx$\to$th interactions are mediated by plastic connections between excitatory populations, while recurrent cx$\to$cx synapses close the excitatory plastic thalamo-cortical loop. During awake states, all neurons have a low value set for the spike-frequency adaptation (SFA) parameter $b$, while during sleep its value is substantially increased for excitatory cortical neurons only.

Throughout awake training and classification, input images are presented to thalamic excitatory neurons encoded as patterns of Poisson stimuli, consequently relayed to the cortex through th$\to$cx connections. In the case of MNIST dataset, images are preprocessed using a histogram of oriented gradients (HOG) filter~\cite{dalal2005histograms}.
Additionally, images from CIFAR-10 are used as an independent dataset; in this case, the preprocessing is implemented via a deep ResNet-like network, designed to extract a set of features that are usable by ThaCo for both awake classification and sleep-driven associative consolidation.
Image preprocessing by HOG or ResNet yields a thalamic input, delivered by the thalamus to the cortex, with an abstraction level compatible to what presented to intermediate layers of the cortex by higher order thalamic nuclei such as the visual pulvinar. In particular, for MNIST, the approach is motivated by observations in primates that the pulvinar receives both cortical (V1/V2) inputs and salience-driven signals from the superior colliculus \cite{shipp2003pulvinar, baldwin2017evolution}, thereby transmitting to V2 a level of representation that already reflects a locally pooled orientation structure, rather than raw retinal signals.

For both MNIST and CIFAR datasets, the network is trained by providing three examples per image class from \num{10} classes, and then evaluated on the task of classifying a balanced test set of \num{250} examples. All the statistics reported are collected over \num{100} trials, randomized across both training and test examples.

During training, contextual signals target example-specific neural assemblies in the cortex to emulate the apical amplification mechanism (Fig.~\ref{fig:net_architecture}B); by contrast, classification is instead performed by switching off the contextual signal (see Secs.~\nameref{sec:methods:training} and~\nameref{sec:methods:classification}).
%
The apical isolation mechanism (Fig.~\ref{fig:net_architecture}C) is implemented during the deep-sleep-like stage by suppressing contextual inputs and, in addition, by increasing SFA and reducing cortical inhibition (see Sec.~\nameref{sec:methods:sleep}). The endogenous deep-sleep-like dynamics is induced by injecting aspecific cortical noise, leading to SO cortical activity patterns that promote thalamo-cortical replay. As anticipated, a significant enhancement of the present ThaCo model compared to its ancestors \cite{capone2019sleeplike, golosio2021thalamocortical} is that cx$\to$th and th$\to$cx excitatory synapses are active and plastic during sleep. This mechanism supports memory consolidation across the entire network through both homeostatic and associative plasticity.

\subsection{Sleep-induced homeostatis}

\begin{figure}[!b]
    \centering
    \includegraphics[width=\textwidth]{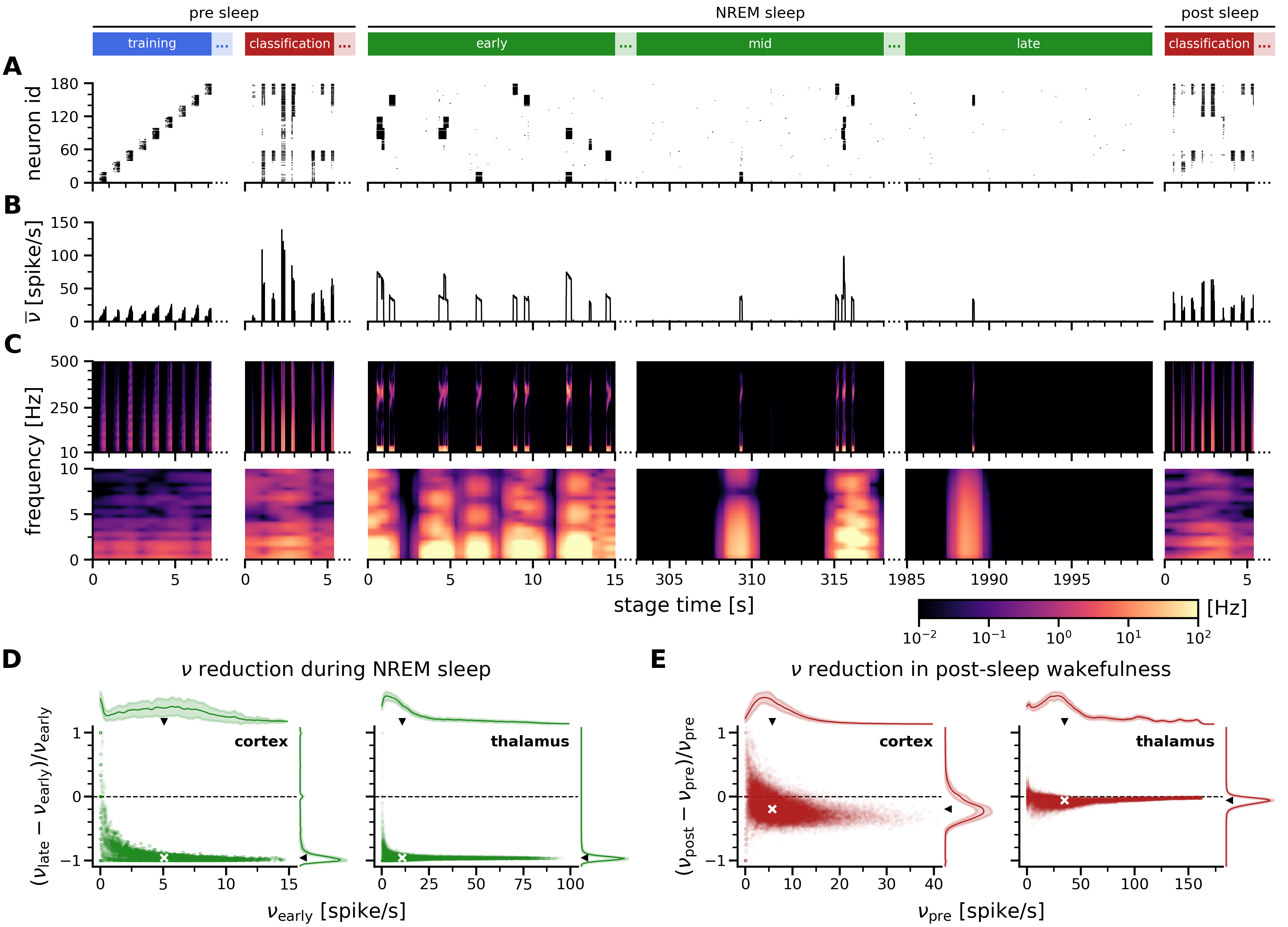}
    \caption{
        \textbf{Network Activity}.
        Evolution of firing rates and spectral content of the thalamo-cortical activity during an awake-sleep learning cycle (training, blue; awake classification, red; NREM sleep, green). The endogenous homeostatic depression happening during sleep affects post-sleep firing rates and spectral content.
        For legibility, a restricted subset of \num{180} neurons is shown in panels A, B, and C, corresponding to the cortical assemblies encoding the first \num{9} learned training examples (representing only the first three classes).
        \textbf{A}) Rastergram of cortical activity.
        \textbf{B}) Average instantaneous firing rate $\overline{\nu}$, obtained from the cortical rastergram in panel A) via a $\sigma=\qty{1.2}{\ms}$ Gaussian convolution.
        \textbf{C}) Corresponding spectrogram, obtained via a narrower ($\sigma=\qty{0.3}{\ms}$) Gaussian convolution to highlight the high-frequency spectral content. Notably, during NREM sleep, delta-band activity is prominent with high-frequency gamma in coincidence with up states. Early sleep manifests a larger frequency of up states than late sleep.
        \textbf{D}) Firing rate change from early to late NREM for cortex (left) and thalamus (right). Main panels are scatter plots, stacking together all the neurons from the \num{100} independent trials (each dot representing a neuron). x-axis is the early spiking rate, while y-axis is its relative change. Medians along both axes are represented with a white cross. Smaller panels at the top and on the right represent trial-averaged marginal distributions for early firing rates and relative changes (shaded area: one standard deviation). Medians from scatter plots are also reported here as black tips, for comparison.
        \textbf{E}) Same as in D), here comparing changes in the firing rate activity during awake post-sleep classification, with respect to pre-sleep classification.
    }
    \label{fig:net_activity}
\end{figure}

Fig.~\ref{fig:net_activity} illustrates the evolution of the cortical activity -- focusing on a subset of \num{180} cortical neurons, corresponding to the first three classes -- during an initial stage of few-shot training from the MNIST dataset, followed by a pre-sleep classification test, a continuous \qty{2000}{\s}-long deep-sleep phase, and a final post-sleep test. For visualization clarity, only some portions of the various network states are shown: training over the first \num{3} classes (\ie,~\num{9} examples); pre-sleep classification on \num{9} examples; initial and final \qty{15}{\s} of sleep, referred to as \textit{early NREM} and \textit{late NREM} respectively, with a further window of \qty{15}{\s} (\textit{mid NREM}) in between them; post-sleep classification on the same \num{9} examples shown in the pre-sleep test.

Focusing on a representative trial, Fig.~\ref{fig:net_activity}A shows the rastergram for the aforementioned subset of \num{180} cortical neurons, with the corresponding instantaneous firing rate -- obtained by convolving the spike trains with a low-pass Gaussian frequency filter ($\sigma=\qty{1.2}{\ms}$, for visualization purposes) -- reported in Fig.~\ref{fig:net_activity}B. The frequency content through the whole learning-sleep protocol is finally reported in Fig.~\ref{fig:net_activity}C, with a zoom on the low-frequency band (\qtyrange[range-phrase=--]{0.5}{10}{\Hz}) and an overview over higher frequencies (\qtyrange[range-phrase=--]{10}{500}{\Hz}). Neuronal firing rates used for computing these spectrograms have been convolved with a less restrictive low-pass filter ($\sigma=\qty{0.3}{\ms}$) with respect to that used for Fig.~\ref{fig:net_activity}B, to highlight the high-frequency spectral content.

During the awake training phase, the sequential presentation of three examples per class causes the activation of different subsets of cortical neurons; thanks to plasticity, engrams are developed as thalamo-cortical assemblies, supporting the selective recall of similar digits during the following classification. Then, during sleep, noise-induced delta-band oscillations emerge and promote co-activation of assemblies through the thalamo-cortical loop. As later discussed, this elicits feature-specific correlations, allowing for the generalization from specific examples to a more abstract class representation.

A first key feature, already observed in previous models and here crucially retrieved, is the progressive but significative activity reduction during sleep, both in number and duration of up states, driven by LTD plasticity. As it can be quite well observed from the spectrogram, high-frequency gamma bursts appear tightly time-locked to cortical up states, consistent with the characteristic oscillatory structure of deep-sleep NREM. In addition, also the neuron activation during post-sleep classification is less pronounced and more regular with respect to pre-sleep classification, though looking exactly at the same test examples. Cortical and thalamic average firing rates are reported in Table~\ref{tab:delta_energy} for the five temporal segments illustrated in panels~\ref{fig:net_activity}A-C, together with an estimate of the frequency of slow-oscillation occurrence during sleep.

\begin{table}[!b]
    \centering
    \caption{
        \textbf{Firing rates during different brain states, and SO frequency during NREM}.
        For each trial, firing rates $\overline{\nu}$ are obtained from averages over neurons, while SO frequency is estimated as the inverse of down-state duration; averages and SEMs over \num{100} independent trials are then reported.
    }
    \label{tab:delta_energy}
    \begin{tabularx}{\textwidth}{l|C{1}|C{1}C{1}C{1}|C{1}}
        \multicolumn{1}{c}{} &
            \multicolumn{1}{c}{\textbf{awake}} &
            \multicolumn{3}{c}{\textbf{NREM sleep}} &
            \multicolumn{1}{c}{\textbf{awake}} \\
        \quad &
            classification &
            early &
            mid &
            late &
            classification \\
        \quad &
            pre-sleep &
            \qtyrange[range-phrase=--]{0}{100}{\s} &
            \qtyrange[range-phrase=--]{300}{400}{\s} &
            \qtyrange[range-phrase=--]{1900}{2000}{\s} &
            post-sleep \\
        \midrule
        $\overline{\nu}$ cx [\unit{\spike\per\s}] &
            \num{6.86 \pm 0.06} &
            \num{5.07 \pm 0.06} &
            \num{1.24 \pm 0.01} &
            \num{0.191 \pm 0.003} &
            \num{5.34 \pm 0.04} \\
        $\overline{\nu}$ th [\unit{\spike\per\s}] &
            \num{48.58 \pm 0.07} &
            \num{18.70 \pm 0.17} &
            \num{4.87 \pm 0.03} &
            \num{0.750 \pm 0.013} &
            \num{45.96 \pm 0.05}  \\
        SO rate [\unit{\Hz}] &
            -- &
            \num{4.45 \pm 0.05} &
            \num{2.36 \pm 0.04} &
            \num{0.93 \pm 0.04} &
            -- \\
    \end{tabularx}
\end{table}

Going further quantitative about these homeostatic effects on the firing rates, in Fig.~\ref{fig:net_activity}D we look at the reduction in both cortical and thalamic firing rates, collecting data over \num{100} independent trials. For each neuron in each trial, the x-axis reports its firing rate $\nu_{\mathrm{early}}$ at the beginning of the sleep phase, while the y-axis reports the relative change in firing rate between early and late NREM, defined as a $(\nu_{\mathrm{late}}-\nu_{\mathrm{early}})/\nu_{\mathrm{early}}$. This representation enables us to identify how sleep-dependent activity evolves across neurons with different baseline excitabilities. The scatter distribution reveals a global down-shift in firing rates, with most points lying below zero, indicating widespread LTD-driven rate depression. A minority of initially low-firing neurons exhibit modest potentiation, consistent with their occasional recruitment during up-state initiation. Trial-averaged marginal distributions (smaller panels on right and top; mean and standard deviation reported) show that both cortical and thalamic populations exhibit the same qualitative trend.

Analogously, also the sleep effects on firing rates during wakefulness can be quantitatively analyzed by comparing post-sleep and pre-sleep classification stages, see Fig.~\ref{fig:net_activity}E. Axes are again defined as initial firing rate $\nu_{\mathrm{pre}}$ (x-axis) vs. relative change $(\nu_{\mathrm{post}}-\nu_{\mathrm{pre}})/\nu_{\mathrm{pre}}$ (y-axis). Here, sleep leads to a pronounced downward shift in cortical firing rates during wakefulness, with thalamic neurons showing a smaller but systematic decrease. These results demonstrate that sleep-induced synaptic reorganization produces state-dependent and long-lasting reductions in network excitability.

\subsection{Cognitive and energetic improvements}

It is already known that the classification accuracy of ThaCo models improves thanks to synaptic reorganization during sleep, leading to generalization and abstraction from examples to classes~\citep{capone2019sleeplike, golosio2021thalamocortical}. Here, we point at quantifying if these cognitive benefits are further enhanced by the extension of plasticity to feed-forward and backward connections between thalamus and cortex, and demonstrate a novel feature: the synaptic reorganization can spontaneously discover multiple hierarchies of categories.

To this aim, we split the \qty{2000}{\s} of deep-sleep in \num{20} chunks of \qty{100}{\s} each, reawakening the network after each epoch of sleep and performing the same classification task, in order to precisely track in time the incremental benefits of sleep. Details about the accuracy definition and this learning-test-sleep-test protocol are reported in Sec.~\nameref{sec:methods}. Results collected over \num{100} independent trials are reported in Fig.~\ref{fig:net_performance}A, comparing the previously published ThaCo configuration with sleep plasticity restricted to cortico--cortical (cx$\to$cx) synapses only (\textcolor{blue}{blue}) \cite{capone2019sleeplike} to the present full-plasticity model, in which thalamo-cortical (th$\to$cx) and cortico--thalamic (cx$\to$th) pathways are additionally plastic during NREM-like activity (\textcolor{red}{red}). For the for the CIFAR-10 case, see Suppl. Fig.~\ref{supp_fig:net_performance_CIFAR}A.
In the MNIST dataset case, enabling both th$\to$cx and cx$\to$cx plasticity during sleep increases accuracy from approximately \qty{\sim 59}{\percent} to \qty{\sim 70}{\percent} (median values over trial), reproducing and outperforming the results obtained in \citet{capone2019sleeplike}, where plasticity limited to cx$\to$cx connections yields a more modest improvement, from \qty{\sim 59}{\percent} to \qty{\sim 66}{\percent}. Corresponding mean and SEM estimates are further detailed in Table~\ref{tab:accuracy}. In the same table, we also report the evolution of the classification performance as a function of the sleep time for the CIFAR-10 dataset case, finding results in the full vs. cx$\to$cx-only plasticity configuration consistent with the MNIST case.

\begin{figure}[!b]
    \centering
    \includegraphics[width=\textwidth]{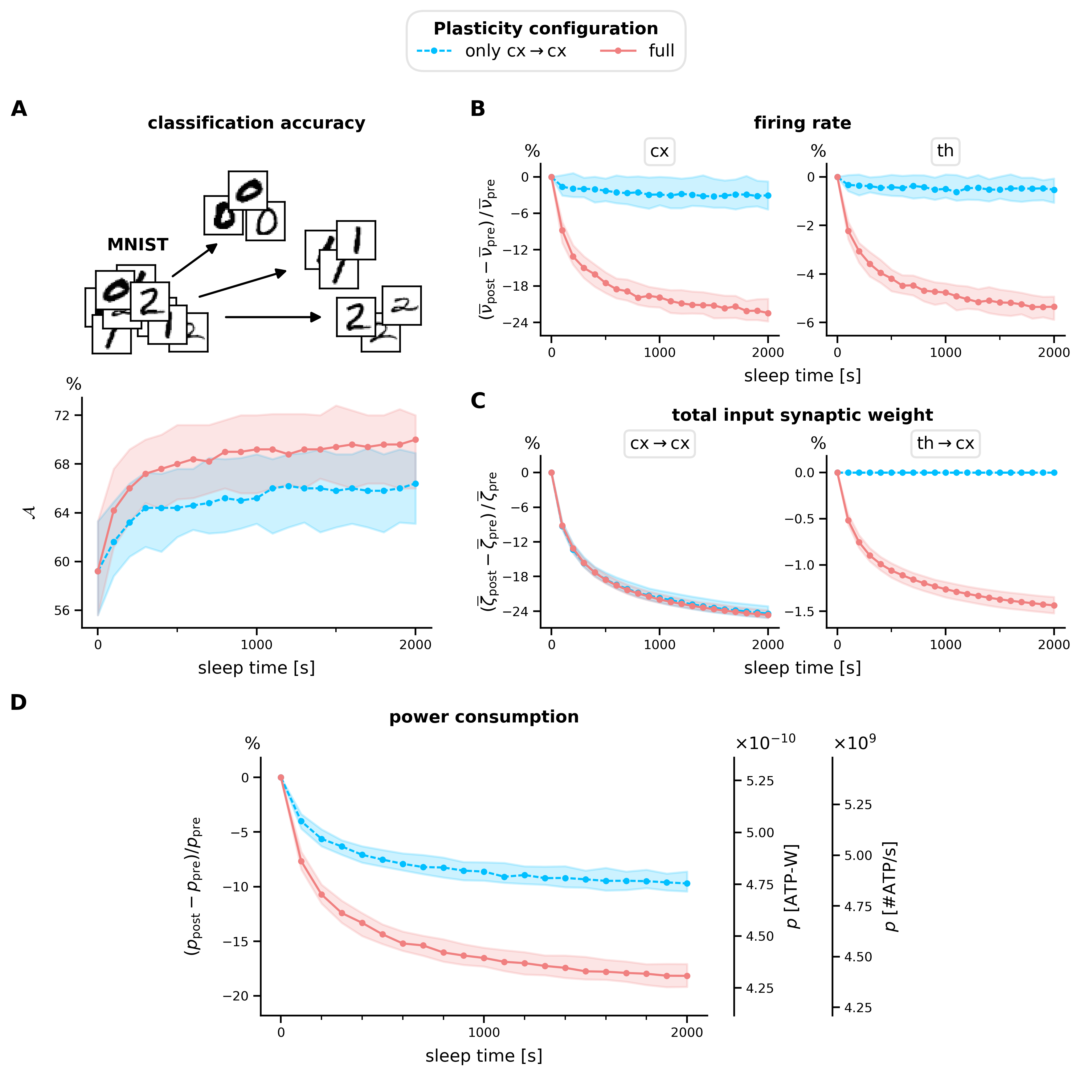}
    \caption{
        \textbf{Sleep effects on cognitive and energetic performance on MNIST classification: comparison with prior ThaCo baseline.} In all panels, \textcolor{blue}{blue} denotes the previously published model~\cite{capone2019sleeplike} with sleep plasticity restricted to cx$\to$cx synapses only, whereas \textcolor{red}{red} denotes the present full-plasticity model (also th$\to$cx and cx$\to$th plasticity are enabled during sleep). Observables are measured post-sleep for incremental sleep duration (\num{20} rounds of \qty{100}{\s}-long NREM-sleep). Network is trained on three MNIST digit examples per class (\num{10} classes in total). Classification is performed over a balanced
        [$\to$ continues on next page]
    }
    \label{fig:net_performance}
\end{figure}

\begin{figure}[!t]
    \centering
    \ContinuedFloat
    \caption*{
        [$\to$ continues from previous page]
        test-set composed of \num{250} images.
        Observables are averaged over \num{100} independent trials: for each of them, median and inter-quartile range are shown. For panels B), C), and D), relative changes with respect to pre-sleep condition are computed, before averaging over trials.
        \textbf{A}) Enhancement of MNIST classification accuracy $\mathcal{A}$.
        \textbf{B}) Reduction of cortical (left) and thalamic (right) single-neuronal firing rate (averaged over neurons from the same layer) $\overline{\nu}$ during classification. In particular, notice the roughly \qty{-20}{\percent} change in the cortical firing rate for the full-plasticity configuration, coherent with the experimental observation in \citet{watson2016network}.
        \textbf{C}) Reduction of total synaptic weights $\overline{\zeta}$ entering cortical neurons from cortex itself (left) and thalamus (right), averaging over all post-synaptic cortical neurons.
        \textbf{D}) Stemming from previous panels, a power consumption ($p$) optimization is incrementally observed during sleep, as quantified at the single-neuron level by Eq.~(\ref{eq:power_single_neuron}). In order to highlight the possibility to directly compare with experiments, we reported relative changes with respect to the pre-sleep value (left y-axis) and absolute values, measured in both \unit{\ATPW} and \unit{\ATP} (right y-axes); see Sec.~\nameref{sec:methods:power} for details. See also the analogous for CIFAR-10 in Suppl. Fig.~\ref{supp_fig:net_performance_CIFAR}.
    }
\end{figure}

By using the same \num{20}-chunk checkpointing technique during sleep, also the evolution of cortical and thalamic firing rates can be tracked, as reported in Fig.~\ref{fig:net_performance}B for the MNIST case, and in Suppl. Fig.~\ref{supp_fig:net_performance_CIFAR}B, for the CIFAR-10 case. Color-coding for the two plasticity configurations is the same as in panel A. Measured as a percentage reduction with respect to pre-sleep classification stage, we find that sleep reduces the post-sleep awake firing rate by approximately \qty{22}{\percent} (cx) and \qty{5}{\percent} (th) (median values over trials), whereas the cortico-cortical-only model shows no significant changes: \qty{\sim 3}{\percent} for cx and less than \qty{1}{\percent} for th.

Another hallmark of activity regularization and optimization is given by total pre-synaptic conductance, measured as the sum of excitatory synapses impinging on cortical neurons from cortex and thalamus, respectively, then averaged over same-region neurons. Fig.~\ref{fig:net_performance}C shows the evolution induced by incremental sleep on these quantities: full-plasticity model exhibits an average reduction of \qty{\sim 25}{\percent} from cx and \qty{\sim 1.5}{\percent} from th; a similar change occurs for the cortical input in the cx$\to$cx-only plasticity configuration (about \qty{-24}{\percent}), while th$\to$cx synapses remain unchanged by construction. Also this result is consistent for the CIFAR-10 dataset, see Suppl. Fig.~\ref{supp_fig:net_performance_CIFAR}C.

\begin{table}[!b]
    \centering
    \caption{
        \textbf{Effect of sleep on classification accuracy on a network trained with only three examples per class}. Performances check-pointed along NREM sleep epochs for the two plasticity configurations (full, vs the cx$\to$cx-only used in~\citet{capone2019sleeplike}) tested on two different datasets (MNIST and CIFAR-10). Mean and SEM values over \num{100} independent trials reported.
    }
    \label{tab:accuracy}
    \begin{tabularx}{\textwidth}{XC{1}C{1}C{1}C{1}C{1}}
        \multicolumn{2}{c}{\multirow{2}{*}{\textbf{Classification accuracy}}} &
            \multirow{2}{*}{Pre-sleep} &
            \multicolumn{3}{c}{Post-sleep} \\
        &
            &
            &
            after \qty{100}{\s} &
            after \qty{400}{\s} &
            after \qty{2000}{\s} \\
        \midrule
        \multirow{2}{*}{MNIST} &
            full plasticity &
            \qty{59.7 \pm 0.5}{\percent} &
            \qty{64.4 \pm 0.5}{\percent} &
            \qty{67.9 \pm 0.4}{\percent} &
            \qty{69.6 \pm 0.4}{\percent} \\
        &
            only cx$\to$cx &
            \qty{59.7 \pm 0.5}{\percent} &
            \qty{62.0 \pm 0.5}{\percent} &
            \qty{64.5 \pm 0.5}{\percent} &
            \qty{66.2 \pm 0.4}{\percent} \\
        \midrule
        \multirow{2}{*}{CIFAR-10} &
            full plasticity &
            \qty{60.4 \pm 0.4}{\percent} &
            \qty{63.4 \pm 0.4}{\percent} &
            \qty{65.0 \pm 0.4}{\percent} &
            \qty{65.9 \pm 0.4}{\percent} \\
        &
            only cx$\to$cx &
            \qty{60.4 \pm 0.4}{\percent} &
            \qty{62.2 \pm 0.4}{\percent} &
            \qty{63.9 \pm 0.4}{\percent} &
            \qty{64.6 \pm 0.4}{\percent} \\
    \end{tabularx}
\end{table}

Sleep-induced decrease in both firing rates and synaptic conductances would naturally lead to an optimized energy consumption for the network. In order to quantify these energetic consequences of sleep-induced synaptic reorganization, we introduce a novel time-dependent biophysically grounded estimator of metabolic power consumption operating at both the single-neuron level, Eq.~(\ref{eq:power_single_neuron}), and at the regional scale, Eq.~(\ref{eq:power_region_full}). The formulation is constrained by key experimental measurements, including PET-derived cortical glucose budgets, ATP metabolic rates, morphometric estimates of synaptic and dendritic membrane fractions, and physiological partitioning of neuronal energy use into housekeeping, spiking, and synaptic components. Full methodological details and experimental groundings are provided in Sec.~\nameref{sec:methods}. At the microscale, the estimator expresses the instantaneous ATP cost of each neuron as the sum of four contributions -- baseline maintenance, memory-maintenance proportional to total synaptic conductance, action-potential generation, and activity-dependent synaptic transmission -- thereby enabling us to track how sleep reshapes cellular-level energy usage, both globally and for each of these four contributions. At the mesoscale, the same formulation aggregates these contributions across cortical or thalamic populations, allowing direct comparison between simulation-derived estimates and experimentally accessible regional measurements of metabolic rates, firing rates, and dendritic and synaptic evolution.

At this point, we are finally able to exactly quantify the energetic benefits of sleep in both ThaCo configurations of full plasticity and cx$\to$cx-only plasticity, as represented in Fig.~\ref{fig:net_performance}D for MNIST. Also in this case, switching on the plasticity on feed-forward and feedback connections allows to further optimize during sleep the network activity, resulting in a reduction of \qty{\sim 18}{\percent} in the estimated metabolic power after sleep, to be compared with the \qty{\sim 10}{\percent} reduction for the reference model with reduced plasticity. Coherent results can be found when looking at the CIFAR-10 dataset, Suppl. Fig.~\ref{supp_fig:net_performance_CIFAR}D.
These findings show that the same mechanisms that reorganize thalamo-cortical representations during NREM also drive a measurable energetic optimization, improving computational efficiency alongside classification performance.

\subsection{Synaptic reorganization}

The global LTD of example-specific synapses induced by sleep, together with a class-specific LTP, can be thoroughly investigated and quantified by leveraging the same \num{20}$\times$\qty{100}{\s} sleep-test checkpoint procedure used so far.

\begin{figure}[!b]
    \centering
    \includegraphics[width=\textwidth]{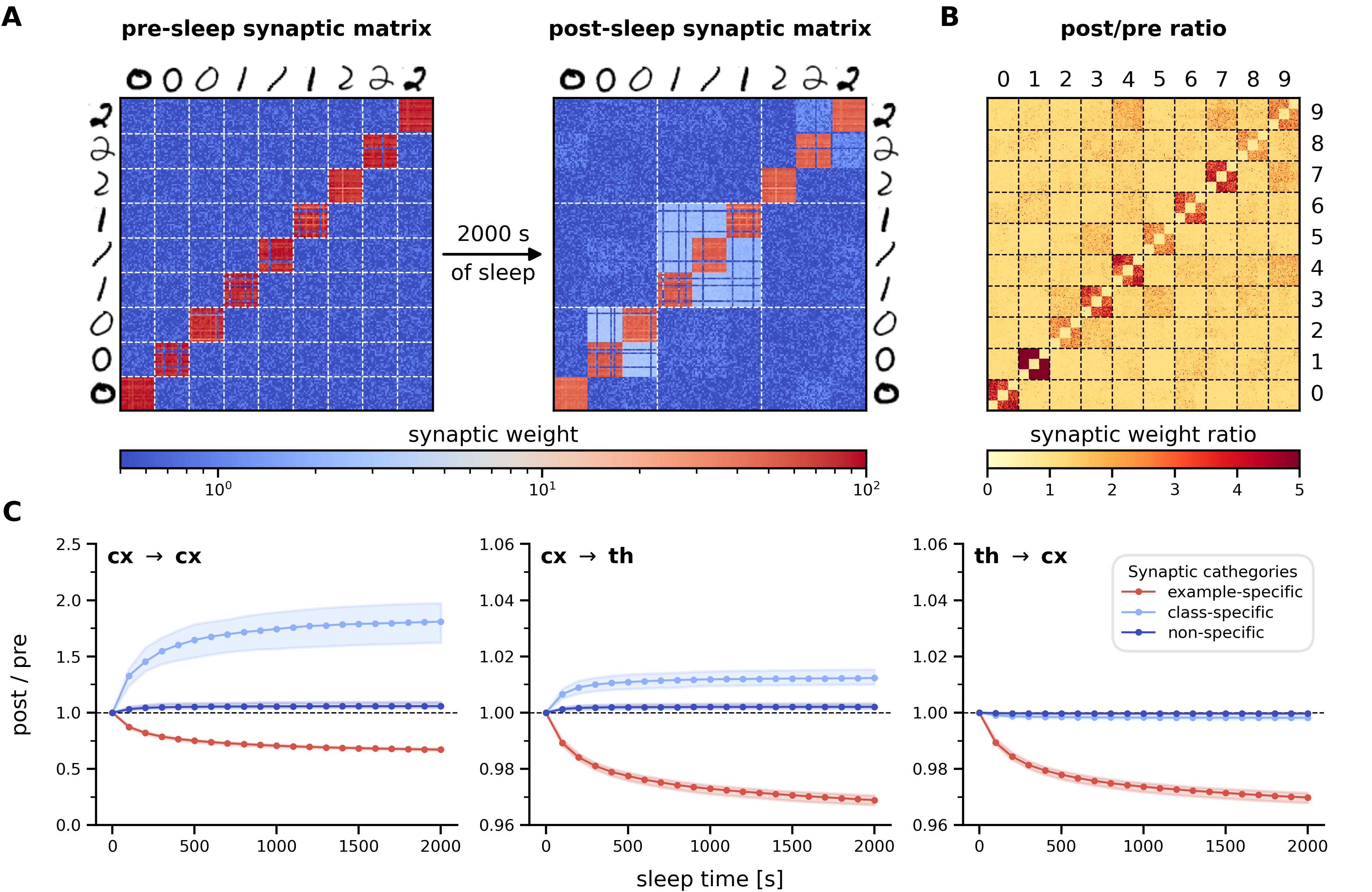}
    \caption{
        \textbf{Network synaptic changes during sleep}.
        During sleep, apical isolation mechanism drives an endogenous dynamics inducing associative and homeostatic effects in synaptic weights.
        \textbf{A}) Sleep effects on the cx$\to$cx synaptic matrix for a representative trial, zoomed on the first three digit classes, three examples per class. Synaptic weight values are color-coded in log scale (reference weight is \qty{1}{\nano\siemens}). Left: before sleep, stronger synapses (red) emerge within example-specific assemblies, sculpted by apical amplification during the initial training, while cross-example synapses (blue) show no sign of association, staying around the randomly initialized values some orders of magnitude below. Right: after \qty{2000}{\s} of sleep, the strength of example-specific synapses reduces (homeostatic depression; from darker to lighter red), while intermediate-strength synapses spontaneously emerge among cell assemblies encoding for the same digit class (memory association; light blue).
        \textbf{B}) Trial-average (\num{100} independent trials) of the entire post-sleep/pre-sleep cx$\to$cx synaptic matrix (ten digit classes), coherently showing the homeostatic and associative effects previously reported for a single trial.
        \textbf{C)} Homeostatic effects for example-specific synapses and associative effects for class-specific synapses shown as a function of sleep time for for the connectivity matrices: cx$\to$cx, cx$\to$th and th$\to$cx; data are computed as the ratio between post-sleep/pre-sleep averages over the three functional classes of connectivity for every given trial; medians and inter-quartile ranges over the means from the \num{100} independent trials are shown. See the analogous for CIFAR-10 in Suppl. Fig.~\ref{supp_fig:net_synapses_CIFAR}.
    }
    \label{fig:net_synapses}
\end{figure}

In Fig.~\ref{fig:net_synapses}A, left side, we show the single-trial synaptic matrix for cortico-cortical connections before entering into the sleep phase -- zooming on the same subset of \num{180} neurons represented in Fig.~\ref{fig:net_activity}A-C. The apical amplification mechanism, taking place during the training stage, promotes the creation of independent cortical assemblies, each encoding individual inputs (strong example-specific synapses). This reflects into the structure of the synaptic matrix, where entries are split in two functional categories according to their magnitude (expressed in log scale): example-specific synapses (strong, color-coded in red, arranged along the main diagonal) and non-specific synapses (weak, color-coded in blue, placed out of the diagonal). Then, after \qty{2000}{\s} of sleep, a remodulation of the synaptic matrix is clearly appreciable, thanks to apical isolation mechanism (Fig.~\ref{fig:net_synapses}A, right side for the first three classes of an exemplary trial, and Fig.~\ref{fig:net_synapses}B for the whole set of classes, averaged over all trials): example-specific synapses are depressed, new ones are formed among neurons initially encoding for different training examples, but belonging to the same class (mid-strength synapses, color-coded in light blue and arranged on the two sides of the main diagonal), while non-specific synapses remain practically unchanged. In other words, the association induced by sleep allows for generalizing from specific examples to abstract classes, and reflects into the appearance of a new class of synapses, the class-specific ones.
Analogous results are reported in Suppl. Fig.~\ref{supp_fig:net_synapses_CIFAR}A for the CIFAR-10 case.

The correlate of the homeostatic and associative effects onto the synaptic matrix is clearly consistent across trials, and is quantified in Fig.~\ref{fig:net_synapses}B by showing as a matrix the post-to-pre sleep synaptic ratio, averaged over trials, with the magnitude of sleep-induced changes color-coded via the corresponding color bar. This time we extend the view to all the cortical neurons, showing that though all input classes are affected by sleep homeostatic and associative effects, the intensity of these effects is clearly inhomogeneous, depending on the similarity between digit features as processed by the HOG preprocessing. At the same time, non-specific synapses remain quite unchanged. Fig.~\ref{fig:net_synapses} is further detailed in Suppl. Table~\ref{supp_tab:synapses}, which summarizes the pre-sleep (post-training) synaptic efficacy values.
Notably, the spontaneous discovery of associations is even more striking in the CIFAR-10 case (see Suppl. Fig.~\ref{supp_fig:net_synapses_CIFAR}B), where several levels of hierarchies are differentially associated during sleep: \textit{i)} at an higher level of abstraction, the distinction among animals and vehicles; \textit{ii)} at an intermediate level, among similar classes (\eg,~between cats and dogs, automobiles and trucks); \textit{iii)} at a lower level, the association between objects belonging to the same image class. Such association between engrams learned during training is non catastrophic, at variance leading to better post-sleep cognitive performance.

Going even more quantitative, and aiming at monitoring the categoric evolution across sleep stages, in Fig.~\ref{fig:net_synapses}C, left, we report the average post/pre ratio for the three cx$\to$cx synaptic functional categories (example-specific, class-specific, and non-specific). Final numerical values after the whole sleep period are reported in Suppl. Table~\ref{supp_tab:synapses}, collecting statistics over the usual \num{100} trials.
Coherently with what observed above, example-specific synapses are systematically depressed via LTD as sleep time increases, while class-specific synapses increase via LTP, and non-specific synapses are only very mildly potentiated.
An even larger class-specific LTP for cx$\to$cx recurrent connections can be observed for the CIFAR-10 dataset, see Suppl. Fig.~\ref{supp_fig:net_synapses_CIFAR}C.
Qualitatively similar results -- though quantitatively less relevant -- can be retrieved in cx$\to$th synapses.
Finally, in feed-forward connections (th$\to$cx), we report the sole homeostatic depression of example-specific connections, while class-specific and non-specific connections stay unchanged on average.

The mechanistic explanation based on the search for similarities in thalamic representation in action during sleep, driven by spontaneous activation of cortical assemblies is supported by Fig.~\ref{fig:preprocessing}D and~\ref{fig:preprocessing}E: they report the correlation in the feature space for both the MNIST and CIFAR-10 cases. For both cases, Fig.~\ref{fig:preprocessing}E shows that the distance between the vectors of features representing examples belonging to the same categories are in average lower than that between examples belonging to different categories, but a notable overlap still persists after the preprocessing, for both the MNIST and CIFAR-10 test cases. The fact that both HOG and ResNet preprocessings have not yet completely separated the examples in categories is also supported by the application of the UMAP learning algorithm \cite{mcinnes2018umap}, as shown by Fig.~\ref{fig:preprocessing}F.

While these changes in the three categories of recurrent connections are not a novelty -- having been already acknowledged in previous ThaCo versions \cite{capone2019sleeplike, golosio2021thalamocortical} -- here the extension of plasticity to connections from and to thalamus allows to address recurrent and backward synaptic changes as the privileged pathways for sleep-induced optimization, with changes in feed-forward connections only linked to homeostatic benefits. 
Furthermore, in the case of the CIFAR-10 dataset, the recurrent inter-layer dynamics is able to create two hierarchical levels of categorization, a novel discovery.

\section{Discussion}
\label{sec:discussion}


ThaCo is a multi-layer spiking network, resembling a thalamo-cortical architecture, capable of learning and classifying visual stimuli. The model exhibits improved classification and energetic performance after a simulated slow-oscillation phase, inspired by biological NREM sleep, which drives the autonomous replay of encoded memories and supports memory consolidation. This is achieved by explicitly allowing all synapses -- feed-forward (thalamo$\to$cortical), feedback (cortico$\to$thalamic), and recurrent (cortico$\to$cortical) -- to be plastic during NREM-like activity. Accordingly, throughout the Results (Fig.~\ref{fig:net_performance}), the intra-layer-limited plasticity configuration (\textcolor{blue}{blue}) provides the direct baseline corresponding to prior ThaCo work \cite{capone2019sleeplike}, while the improvements reported here isolate the contribution of extending sleep plasticity to inter-layer thalamo-cortical and cortico-thalamic pathways (\textcolor{red}{red}).
The resulting endogenous dynamics then induces widespread synaptic modifications through competitive homeostatic-associative mechanisms, driving a global downscaling of excitatory synaptic weights and promoting the unsupervised formation of new synapses among engrams encoding examples belonging to the same input class, and also between similar classes (\eg,~animals vs. vehicles in the the CIFAR-10 dataset case). The simulated sleep dynamics also features the experimental canonical signatures of deep NREM, namely the reduction of the slow-oscillation rate of occurrence, and the firing rate homeostasis.
Additionally, a significant reduction in overall thalamic and cortical firing rates during wakefulness is observed, allowing the network to operate more energy-efficiently after sleep, considerably reducing the metabolic power consumption -- here estimated by introducing a novel biologically-grounded definition. These results suggest that full inter- and intra-layer plasticity during sleep is essential for coordinating the activation of neighbouring layers in a hierarchical network, and for down-regulating cortical excitability during post-sleep wakefulness, leading to an overall task-oriented optimization of the network.

A salient feature of this model is its grounding in biological principles, applied at defining both the network architecture and the mechanisms that support learning and memory reorganization while preventing catastrophic forgetting, conversely producing post-sleep classification improvements.
The model input poses at an abstraction level similar to what encoded by the thalamic visual pulvinar nuclei, therefore constituting a plausible biological analogue for the input to V2 layer (for the MNIST case) or higher areas (for the CIFAR-10 case) in the hierarchical visual cortex architecture.

Few-shot learning is enabled by the apical amplification principle, that is implemented solely through the temporal coincidence of external and internal specific stimuli, supported by a hard winner-take-all mechanism. Then, the apical isolation regime during sleep allows for search of memories with similar representations, supported by a soft winner-take-all mechanism.

Notably, the proposed mechanism is generalizable to any neighbouring pair of network layers, opening interesting paths in the bio-inspired artificial intelligence research.

A second contribution of this work is the introduction of a quantitative methodology to estimate the energy consumption in plastic spiking neural networks, that explicitly accounts for the evolution of synaptic weights, grounding on measures performed at subcellular, cellular and areal scales. 

Even if the estimator -- at its current stage of calibration -- is intended more for comparative analyses within controlled simulation settings, rather than as an absolute prediction of \textit{in vivo} metabolic expenditure, the proposed approach allows for directly comparing energetic estimates from simulations with experimental metabolic data, and could be leveraged for multi-modal data integration, driving both a more accurate calibration of learning models, and a better quantitative understanding of the main factors contributing to metabolic consumption in biological brains engaged in learning.

To date, few models have been developed to investigate the effects of plastic sleep on synaptic structure and task-related performance. Despite some conceptual similarities, existing models generally lack endogenous sleep dynamics that can spontaneously give rise to synaptic homeostasis, associative plasticity, and firing rate downscaling, in fact typically relying on \textit{ad hoc} regularizations. These dynamics are here achieved in an unsupervised manner using a pairwise STDP rule, without requiring explicit regularization terms.

Despite the solid conceptual framework of the model, a few limitations remain.
First of all, a proper implementation of the apical amplification mechanism would require at least a two-compartment neuron, capable of generating short bursts of spikes in response to the temporal coincidence of perceptual and contextual stimuli~\cite{larkum2013cellular, capone2022burst, capone2023beyond, pastorelli2025simplified}, paving the way toward dendritic computation at the single-neuron level~\cite{poirazi2020illuminating, chavlis2025dendrites}.
In addition, the model does not include mechanisms for plasticity during awake classification, nor does it account for REM sleep, day-dreaming, or resting-state activity.

Aiming at addressing the aforementioned limitations, in future works we would like to address \textit{i)} the incorporation of a biologically grounded two-compartment neuron \cite{pastorelli2025simplified}, and \textit{ii)} the investigation of its capacity for rapid incremental learning with biologically plausible firing rates, extending results from \citet{golosio2021thalamocortical}.
In addition, \textit{iii)} we plan to further extend the ThaCo architecture, developing a multi-area network capable of assimilating a multimodal input, thereby enabling the study of inter-areal associative mechanisms.
In parallel, \textit{iv)} we intend to model and implement a REM-sleep-like phase, and the related required variant of apical mechanisms (namely, the apical drive). This is essential to integrate information across multiple cortical areas, a task that cannot be performed during NREM, due to the apical-isolation mechanism in action, that limits synaptic reorganization to local intra- and inter-layer circuits. The simulation of REM will enable the network to optimize its entire synaptic structure, along the line prototyped by some of the authors of this paper in \citet{deluca2023nrem}.

\section{Methods}
\label{sec:methods}

\subsection{ATP-calibrated metabolic estimator for plastic spiking networks}
\label{sec:methods:power}

Adenosine triphosphate (ATP) is the cellular energy currency that supports neurotransmission, biosynthesis and recycling of neurotransmitters, and membrane potential dynamics; in turn, ATP is synthesized through oxidative phosphorylation of adenosine diphosphate (ADP) using glucose as the primary fuel~\cite{mckenna2012energy, jamadar2025metabolic}.

According to recent positron emission tomography (PET) studies summarized in~\citet{levy2021communication}, the total metabolic power potentially available to the human brain through glucose oxidation is \qty{17.0}{\W}. Of this, about \qty{11}{\percent} is actually not oxidized, and \qty{8.89}{\W} is dissipated as heat, leaving roughly \qty{6.19}{\W} available for biochemical work from ATP at the whole-brain level; to stress this, from now we will use the notation \unit{\ATPW}, rather than simply \unit{\W}. Regional partitioning assigns about \qty{3.09}{\ATPW} to cortical gray matter, \qty{1.85}{\ATPW} to white matter, and the rest to cerebellum and remaining brain regions.

The energy consumption of neurons in cortical gray matter is usually considered to be the sum of contributions required by a set of main subcellular processes: \textit{i)}~resting potential maintenance, \textit{ii)}~action potential production, \textit{iii)}~post-synaptic activity and receptor rebalancing, \textit{iv)}~neurotransmitter recycling, \textit{v)}~pre-synaptic vesicle management, and \textit{vi)}~other non-signaling housekeeping tasks (see \citet{howarth2012updated, mckenna2012energy}, updated by \citet{engl2015nonsignalling}). In the same references, they also provide estimates of the relative and total energetic costs of the different sub-cellular processes for a cortex firing on average at \qty{4}{\Hz}, partitioning the energy budget into three main categories: \textit{1)~housekeeping and resting potential} (\qty{\sim 40}{\percent} of the energy income); \textit{2)~action potentials} (\qty{\sim 16}{\percent}); \textit{3)~synaptic transmission} (\qty{\sim 44}{\percent}). Only glutamatergic neurons are considered in these estimates, based on their numerical and synaptic dominance (\qty{\sim 95}{\percent} of projections, according to \citet{dekock2023shared}). An experimental \textit{in vivo} technique able to assess the cerebral ATP metabolic rates is $^{31}$P-MT, a magnetic resonance spectroscopy (MRS) combined with the magnetization transfer (MT) method; $^{31}$P-MT performed under increasing anesthesia levels~\cite{du2008tightly} -- up to silent, isoelectric states -- indicates both a tight coupling between ATP metabolic rates and brain activity, and the maintenance of a substantial baseline in the absence of spiking activity accounting for at least \qty{40}{\percent} of the ATP consumption measured under mild anesthesia, confirming the previous baseline estimates. 

Aiming at the a compact power estimator for plastic spiking networks, we building on these established measurements, and propose the definition of a distinct \textit{memory-maintenance} subcategory within the broader \textit{housekeeping and resting potential} energy budget, encompassing the metabolic processes required to preserve synaptic structure and molecular memory traces.
This leads to the formulation of a novel power-consumption estimator at the single-neuron level:
\begin{equation}
    p_i(t) = p_i \bigl[\boldsymbol{w}_i(t), \boldsymbol{\nu}(t)\bigr] = b_0 + b_1 \, \sum_j w_{j \to i}(t) + b_2 \, \nu_i(t) + b_3 \, \sum_j \nu_j(t) \, w_{j \to i}(t) \, ,
    \label{eq:power_single_neuron}
\end{equation}
where $\boldsymbol{w}_i$ is the set of presynaptic weights for neuron $i$, indexed by the source neuron~$j$, and $\boldsymbol{\nu}$ is the set of all presynaptic firing rates, included that of neuron $i$ itself. In the following, we will briefly refer to the sum of all synaptic weights that enter neuron $i$ as the total input conductance $\zeta_i \equiv \sum_j w_{j \to i}$. The \num{4}-parameter set $\boldsymbol{b}$ will be fixed later in this section using the discussed experimental considerations. Therefore, Eq.~(\ref{eq:power_single_neuron}) provides a quantitative approach to accurately estimate the $p_i$ metabolic cost of neuronal activity in simulations.

It is important to notice the explicit time-dependence of Eq.~(\ref{eq:power_single_neuron}), that stresses the fact that biological and artificial spiking networks can change their dynamical state and optimize their working point in both the firing rate and synaptic conductance, eventually resulting in both cognitive and energetic beneficial effects, as discussed in Sec.~\nameref{sec:results}.

At the regional level, \eg~for cortical gray matter, one should sum over all the single-neuron contributions involved, taking into account in the second and fourth term also the contributions of presynaptic neurons potentially coming from other regions. This leads to the following coarse-grained regional power-consumption estimator:
\begin{equation}
    \begin{split}
        P_{\mathrm{r}}(t) = \sum_{i \in \mathrm{r}} p_i(t) &= \sum_{i \in \mathrm{r}} \Bigl( b_0 + b_1 \, \zeta_i(t) + b_2 \, \nu_i(t) +  b_3 \, \sum_j \nu_j(t) w_{j \to i}(t) \Bigr)\\
        &= N_{\mathrm{r}}\Bigl( b_0 + b_1 \overline{\zeta}_{\mathrm{r}}(t) + b_2 \, \overline{\nu}_{\mathrm{r}}(t) +  b_3 \, \overline{\iota}_{\mathrm{r}}(t) \Bigr)\, ,
    \end{split}
    \label{eq:power_region_full}
\end{equation}
where we introduced the time-dependent averages over the $N_{r}$ neurons in the region: $\overline{\zeta}_{\mathrm{r}}(t) \equiv \sum_i \zeta_i(t) / N_{\mathrm{r}}$ for the total presynaptic conductance, $\overline{\nu}_{\mathrm{r}}(t) \equiv \sum_i \nu_i(t) / N_{\mathrm{r}}$ for the firing rate, and $\overline{\iota}_{\mathrm{r}}(t) \equiv \sum_i \sum_j \nu_j(t) w_{j\to i}(t) / N_{\mathrm{r}}$ for the presynaptic input.



This regional estimate has a two-fold relevance: on one side, it enables a quantitative estimation of the energetic effects of synaptic evolution during the simulation, \eg~after a sleep period; on the other side, it enables the comparison with large-scale experimental metabolic data, by grouping the contributions from all the neurons in the region to the different sub-cellular processes into four categories, here below discussed.

The first term, $b_0\,N_{\mathrm{r}}$, represents the regional \textit{baseline consumption}, unrelated to memory maintenance: the synapse-independent cost of maintaining cellular integrity, ion gradients, resting potential, and organellar turnover in the soma and dendritic shafts.

The second term, $b_1\,N_{\mathrm{r}}\,\overline{\zeta}_{\mathrm{r}}(t)$, that depends on the total input synaptic conductance, represents the \textit{memory maintenance cost} associated with the preservation of the learned synaptic inventory at rest -- that is, the energy required to keep existing synapses (spines and boutons) alive and functionally stable after learning. Taken together, these two terms account for the \qty{\sim 40}{\percent} of the energy income previously associated to the subcellular contribution \textit{1) housekeeping and resting potential}; later in this section, we will discuss how to estimate their relative weight.

The third term, proportional to the regional average firing rate $b_2\,N_{\mathrm{r}}\,\overline{\nu}_{\mathrm{r}}(t)$, accounts for the production of action potentials and their axonal propagation to all synapses, and accounts for the aforementioned \qty{\sim 16}{\percent} of the total energy cost associated to \textit{2) action potentials}.

Finally, the fourth term, involving the average over the total presynaptic activity $b_3\,N_{\mathrm{r}}\,\overline{\iota}_{\mathrm{r}}(t)$ of neurons in the region, represents activity-dependent synaptic activation, postsynaptic current injection, neurotransmitter release, and short-term recycling (\qty{\sim 44}{\percent} of the budget, as reported by experiment measures to be associate with point \textit{3) synaptic transmission}).

When interested at the optimization of the power in a given region, for example during a training or a sleep period, a convenient estimate for comparison with experiments can be obtained introducing two approximations in Eq.~(\ref{eq:power_region_full}). The first assumes that the average firing rate of remote regions is the same as the local one. In the second, the full expression of synaptic activity in the fourth term is decoupled as the product of the average firing rate times the average incoming conductance, $\overline{\iota}_{\mathrm{r}}(t) \simeq \overline{\nu}_{\mathrm{r}}(t) \, \overline{\zeta}_{\mathrm{r}}(t)$, neglecting the unavoidable correlations between the firing rate of a presynaptic neuron and the related presynaptic weight. Under these simplifying assumptions, $P_{\mathrm{r}}$ becomes a function of only two scalar region-specific dynamical variables, the regional average synaptic conductance~$\overline{\zeta}_{\mathrm{r}}(t)$ and the regional average firing rate $\overline{\nu}_{\mathrm{r}}(t)$, getting the following expression:
\begin{equation}
    P^*_{\mathrm{r}}(t) = P^*_{\mathrm{r}}\Bigl[\overline{\zeta}_{\mathrm{r}}(t), \overline{\nu}_{\mathrm{r}}(t)\Bigr]
    = N_{\mathrm{r}}\Bigl( b_0 + b_1 \, \overline{\zeta}_{\mathrm{r}}(t) + b_2 \, \overline{\nu}_{\mathrm{r}}(t) + b_3 \, \overline{\zeta}_{\mathrm{r}}(t) \, \overline{\nu}_{\mathrm{r}}(t)\Bigr) \, .
    \label{eq:power_region_proxy}
\end{equation}
Notably, this simplified version enables a direct comparison with experimental measures acquired only at a regional scale.

At this point, we can estimate the values of the four $\boldsymbol{b}$ coefficients from existing experimental data for cortical gray matter, leveraging the $P^*_{\mathrm{r}}(t)$ approximating function and guided by morphometric reconstructions of cortical pyramidal neurons.
Notice that, in general, specific values depend on neuron type, cortical area, and axon arborization geometry; however, our aim here, as modelers, is to provide the community with a starting set of values for the $\boldsymbol{b}$ metabolic coefficients and a convenient tool, rather than to fix their exact values -- a task that still requires substantial experimental and atlasing integration efforts, going beyond the scope of this work. Also, notice that in what follows, uncertainties over experimental measures are reported as \textit{SEM} errors, unless otherwise specified.

We start by computing the fraction of neuron membrane dedicated to memory maintenance (the sum of boutons and spines) and that of dendritic shafts supporting them, compared to which the somatic area is negligible. Our assumption is that this way it is possible to estimate the relative costs of memory maintenance and baseline. Each cortical pyramidal neuron in human L2/3 is estimated to host at least \num{3.0e4} distinct synapses~\cite{eyal2018human}. In fact, each connected pair of neurons is usually supported by multiple synapses: an average of \num{4.0 \pm 0.5} pre-synaptic boutons per excitatory connection is measured in human temporal cortex \cite{hunt2022strong}, this way supporting at least \num{\sim 7.5e3} connections per neuron. In human L5, each pre-synaptic bouton has an average surface area of approximately \qty{6.0 \pm 0.5}{\micro\m^{2}} and each bouton touches an average of \qty{1.03 \pm 0.03} post-synaptic spines~\cite{yakoubi2018quantitative}. By leveraging the \textit{F-factor} metric reported in \citet{hunt2022strong} --  defined as $F=(\text{spine area} + \text{shaft area})/(\text{shaft area})$ -- with a lower estimate for the area of dendritic shaft provided by \citet{rich2020modeling}, we can estimate the percentage of neuronal membrane surface dedicated to bouton and spines -- and hence to memory maintenance -- to be around \qty{89}{\percent}. This finally implies that the \qty{\sim 40}{\percent} associated with the initial \textit{housekeeping and resting potential} contribution can be split into a \qty{\sim 4}{\percent} for the actual \textit{baseline consumption} $b_0 \, N_{\mathrm{gray}}$, and a \qty{\sim 36}{\percent} for the \textit{memory maintenance cost} $b_1 \, N_{\mathrm{gray}}\, \overline{\zeta}_{\mathrm{gray}}$.

The number $N_{\mathrm{gray}}$ of cortical neurons can be estimated to amount to \num{1.63e10} \cite{azevedo2009equal, yang2025principle}, while $P_{\mathrm{gray}} = \qty{3.09}{\ATPW}$ \cite{levy2021communication}, as already stated, then resulting in a metabolic rate per neuron equal to:
\begin{equation}
    p_{\mathrm{gray}} = \qty{1.90e-10}{\ATPW\per\neuron} \, .
    \label{eq:power_single_neuron_measure_W}
\end{equation}
In addition, the energy of ATP cytosolic phosphorylation in brain tissue has been measured~\citep{veech1979cytosolic} as amounting to $|\Delta G_{\mathrm{ATP}}| = \qty{14.08 \pm 0.01}{\kilo\cal\per\mol} = \qty{9.78e-20}{\J\per\ATP}$, then such energy consumption rate per neuron can be also expressed in terms of ATP molecules number, $p_{\mathrm{gray}} = \qty{1.94e9}{\ATP\per\s\per\neuron}$.

Assuming the reference firing rate $\overline{\nu}_{\mathrm{gray}}=\qty{4}{\Hz}$ used by \citet{howarth2012updated, engl2015nonsignalling} for estimates of energetic measures in cortical gray matter and remote regions from which external presynapses originate, the actual values for $\boldsymbol{b}$ coefficients in cortical gray matter can be eventually computed:
\begin{equation}
    \begin{split}
        b_0 = 0.04 \, &p_{\mathrm{gray}} \, , \quad b_1 = 0.36 \, p_{\mathrm{gray}} / \overline{\zeta}_{\mathrm{gray}}, \quad b_2 = 0.16 \, p_{\mathrm{gray}} / \overline{\nu}_{\mathrm{gray}} \, ,\\
        &b_3 = 0.44 \, p_{\mathrm{gray}} / \overline{\iota}_{\mathrm{gray}} \simeq 0.44 \, p_{\mathrm{gray}} / (\overline{\nu}_{\mathrm{gray}} \, \overline{\zeta}_{\mathrm{gray}}) \, .
    \end{split}
\end{equation}
For our network, we get numerical values reported in Table~\ref{tab:b_coefficients}, both in terms of \unit{\ATPW} and number of ATP molecules per second. To this aim, we used the trial-averaged initial (\ie,~pre-sleep) value of total synaptic conductance $\overline{\zeta}_{\mathrm{gray}} = \qty{2.05e-6}{\siemens}$. Notice that the proxy for $b_3$ has been used only for its estimation from experimental data; in all our power measures in simulations, as reported in Sec.~\nameref{sec:results}, we used the non-approximated expressions Eqs.~(\ref{eq:power_single_neuron}) and~(\ref{eq:power_region_full}).

\begin{table}[t]
    \centering
    \caption{
        \textbf{$\boldsymbol{b}$ coefficients for the ThaCo model analyzed in this paper.} Values are computed as averages over the \num{100} (``full plasticity'') plus \num{100} (``cx$\to$cx-only'') independent trials (initialized in a coherent way, so pre-sleep values of $\overline{\zeta}_{\mathrm{gray}}$ can be safely compared) for the MNIST dataset.
    }
    \label{tab:b_coefficients}
    \begin{tabularx}{\textwidth}{L{0.3}C{1.35}C{1.35}}
        &
            \textbf{value in \unit{\ATPW}} &
            \textbf{value in \unit{\ATP\per\second} molecules} \\
        \midrule
        $b_0$ &
            \qty{7.58e-12}{\ATPW\per\neuron} &
            \qty{7.75e7}{\ATP\per\s} \\
        $b_1$ &
            \qty{3.33e-5}{\ATPW\,\Omega\per\neuron} &
            \qty{3.40e14}{\ATP\,\Omega\per(\s\,\neuron)} \\
        $b_2$ &
            \qty{7.58e-12}{\ATPW\s\per\neuron} &
            \qty{7.75e7}{\ATP\per\neuron} \\
        $b_3$ &
            \qty{1.02e-5}{\ATPW\,\Omega\,\s\per\neuron} &
            \qty{1.04e14}{\ATP\,\Omega\per\neuron} \\
    \end{tabularx}
\end{table}

The power consumption estimate here presented has a few limitations, starting from the still fragmentary morphometric information. First, it is based on consideration that focus on excitatory neurons only, an approximation partially justified by the fact that about \qty{95}{\percent} of synapses in the cortex are excitatory \cite{dekock2023shared}.  Second, the partition of energy between neurons and glia in gray matter is currently assumed \citep{howarth2012updated} to be \qty{\sim 75}{\percent} and \qty{\sim 25}{\percent}, respectively, but we still follow the conventional choice of attributing the glial consumption to the neurons they support, as in the references cited above. Third, we based the estimate of $\boldsymbol{b}$ metabolic constants on experimental studies that fixed the firing rate at an average reference value of \qty{4}{\Hz}, while our simulations operate at slightly higher frequencies, spanning from \qty{\sim 6.9}{\Hz} before sleep to \qty{\sim 5.3}{\Hz} at the end of sleep for the cortical region. Fourth, we ignored in the estimate of $b_3$ the correlations between the firing rate of each presynaptic neuron and its presynaptic weight. Finally, depending on the modelling scale, the set of $\boldsymbol{b}$ constants can include a \textit{white matter} cost (\ie,~the long-range axonal transport of spikes through myelinated fibers), which may be taken as proportional to the total axonal arborization length; therefore, the inclusion of white-matter contributions would further shift the balance toward communication costs.

\subsection{Input to the neural network}
\label{sec:methods:input}

\begin{figure}[!b]
    \centering
    \includegraphics[width=\textwidth]{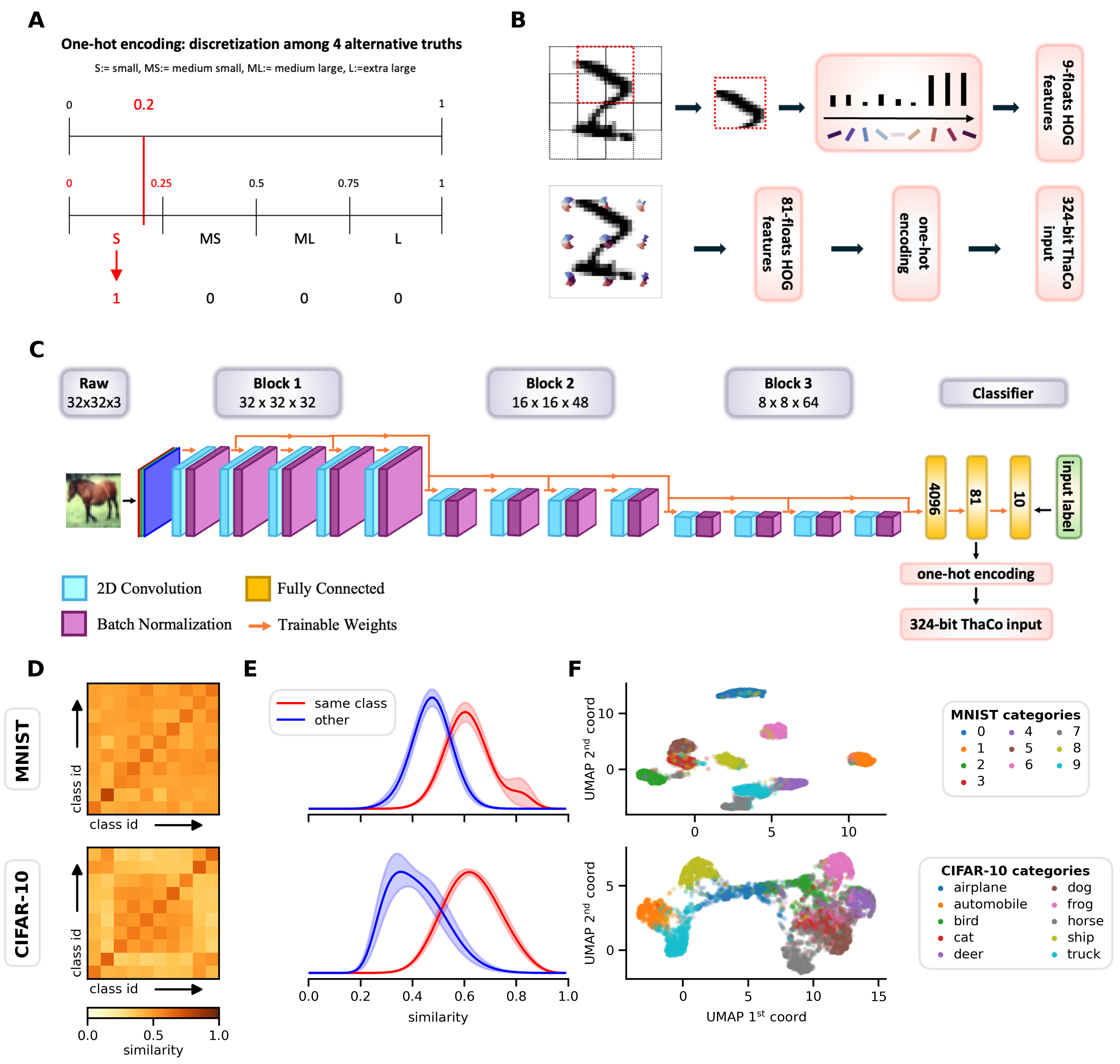}
    \caption{
        \textbf{MNIST and CIFAR-10 input preprocessings and resulting structures}. Raw images drawn from MNIST and CIFAR-10 datasets are processed into \num{81} float values using HOG filters and CNN algorithms, respectively. 
        \textbf{A}) Each input feature (\num{81} floating-point values in range \numrange[range-phrase=--]{0}{1}) is one-hot encoded using a four-level discretization scheme (\numrange[range-phrase=--]{0}{0.25}, \numrange[range-phrase=--]{0.25}{0.50}, \numrange[range-phrase=--]{0.50}{0.75}, \numrange[range-phrase=--]{0.75}{1}) (adapted from \citet{golosio2021thalamocortical}).
        \textbf{B}) Histogram of oriented gradients computed from \num{28}$\times$\num{28} MNIST images using \num{9} overlapping \num{7}$\times$\num{7} kernels. \num{81}-floats resulting features are expanded by the one-hot encoding digitization to produce the \num{324}-bit input to ThaCo.
        \textbf{C}) Residual convolutional neural network (ResNet)
        [$\to$ continues on next page]
    }
    \label{fig:preprocessing}
\end{figure}

\begin{figure}[!t]
    \centering
    \ContinuedFloat
    \caption*{
        [$\to$ continues from previous page]
        preprocessor for CIFAR-10. Feed-forward convolutional layers (light blue), batch normalization (violet), organized into \num{3} blocks of increasing depth and decreasing size. Orange arrows: single layer feed-forward trainable weights. The \num{4096} features are then projected to \num{81} floats, from which the one-hot encoding produces the \num{324}-bit ThaCo input.
        \textbf{D}) Class similarity matrices, quantified as a normalized dot product between the \num{324}-bit feature vectors, averaged over the whole MNIST and CIFAR-10 training and test datasets. In addition to the diagonal category specific structure observed for MNIST, notice the block structure observed in CIFAR-10, separating animal from vehicle image categories, and further additional similarities (\eg,~between cats and dogs); both emerge spontaneously during sleep-dependent processing in ThaCo (Fig.~\ref{fig:net_synapses}B and Suppl. Fig.~\ref{supp_fig:net_synapses_CIFAR}B).
        \textbf{E}) Distributions of intra-class (blue) and inter-class (red) similarities computed over the whole datasets.
        \textbf{F}) UMAP dimensionality reduction \cite{mcinnes2018umap} projects each example from the whole preprocessed dataset in a two-dimensional space, highlighting the partial split in categories.
    }
\end{figure}

The network input comprises training and test images independently drawn for each of the \num{100} trials. Two categorical datasets are used in input: \textit{i)}~the MNIST dataset~\cite{deng2012mnist} (a collection of \qty{60}{\kilo\relax} \num{28}$\times$\num{28}-pixel grayscale handwritten \numrange[range-phrase=--]{0}{9} digit images used as training set, plus \qty{10}{\kilo\relax} test images), and \textit{ii)}~the CIFAR-10 dataset~\cite{krizhevsky2009cifar} (\num{32}$\times$\num{32}-pixel RGB-coloured training images, divided into \qty{50}{\kilo\relax} training and \qty{10}{\kilo\relax} test images, and organized into ten categories -- ``airplane'', ``automobile'', ``bird'', ``cat'', ``deer'', ``dog'', ``frog'', ``horse'', ``ship'', and ``truck'' -- qualitatively expected to be arranged into hierarchies of similarities).
For both datasets, inputs are preprocessed -- see dataset-specific procedure below -- to extract a set of \num{81} features. Each feature is then discretized according to a four-level scheme and subsequently one-hot encoded into a \num{324}-bit binary vector (see Fig.~\ref{fig:preprocessing}A for details), following the procedure introduced in \citet{capone2019sleeplike} and further discussed in \citet{golosio2021thalamocortical}. These binary vectors determine the subsets of thalamic neurons recruited during image presentation in the awake state and reactivated during sleep.

\textbf{MNIST preprocessing.} \quad MNIST digits are preprocessed using a \textit{histogram of oriented gradients} (HOG) filter \cite{dalal2005histograms}. This step extracts edge orientation information while preserving the spatial structure of the features (Fig.~\ref{fig:preprocessing}B). As discussed in Sec.~\nameref{sec:results:setup}, such features express a level of abstraction similar to what conveyed from V1 to V2 by the thalamic pulvinar nucleus.

\textbf{CIFAR-10 preprocessing.} \quad As discussed in Sec.~\nameref{sec:results:setup}, we bring also the features of CIFAR-10 at a level of abstraction similar to what projected by the pulvinar to higher-order visual areas. To this purpose, images are preprocessed using a convolutional neural network (CNN) based on a deep residual architecture \cite{he2016residual}. As illustrated in Fig.~\ref{fig:preprocessing}C, the model employs \num{13} convolutional layers with batch normalization organized into \num{3} blocks of increasing depth (\num{32}, \num{48}, and \num{64}, respectively) and decreasing spatial resolution (\num{32}$\times$\num{32}, \num{16}$\times$\num{16}, and \num{8}$\times$\num{8}, respectively). Through this hierarchical structure, the number of input channels (\num{3}, corresponding to the RGB encoding) is gradually expanded to train suitable filters for extracting meaningful features, while input spatial resolution (\num{32}$\times$\num{32}) is progressively reduced to compress information and promote the emergence of high-level abstract representations. Residual  connections introduce shortcuts between non-adjacent layers, mitigating gradient vanishing during back-propagation. The final convolutional layer is flattened and projected onto an \num{81}-unit linear layer, which, after the one-hot encoding, yields the feature vector provided as input to ThaCo.

Figs.~\ref{fig:preprocessing}D-F provide additional information about the structure of the input features produced by the HOG and ResNet preprocessors for the MNIST and CIFAR-10 datasets, respectively. Notably, a clear similarity among CIFAR-10 features manifests not only between examples belonging to the same category, but also differentially grouping examples of animals and inanimate objects, and between more similar categories (\eg,~between cats and dogs).

\subsection{Network architecture}
\label{sec:methods:architecture}

The model consists of a two-layer spiking neural network designed to abstract thalamo-cortical interactions (Fig.~\ref{fig:net_architecture}D). Each layer is composed of interconnected excitatory and inhibitory populations made up of conductance-based leaky-integrate-and-fire (LIF) adaptive exponential (AdEx) neurons with spike-frequency adaptation \citep{gerstner2014neuronal}.

At the lowest hierarchical level, the thalamus is composed of \num{324} excitatory neurons, corresponding to the length of the input pattern, and \num{200} inhibitory neurons. The perceptual input, encoded according to the one-hot-encoding scheme (see Fig.~\ref{fig:preprocessing}A), drives thalamic excitatory neurons, by means of a set of Poisson processes with rates $\nu_{\mathrm{th},\,0}$, $\nu_{\mathrm{th},\,1}$. Thalamic excitatory neurons projects to all excitatory cortical neurons, through bidirectional plastic connections. No recurrent connections between thalamic neurons are included.

A second layer, representing the cortex, consists of \num{600} excitatory and \num{200} inhibitory neurons. The cortical excitatory population exhibits recurrent plastic connections and, during the training phase, receives in addition to the thalamic input (perceptual signal) a Poisson stimulus emulating a time-dependent contextual modulation input from other brain areas (contextual signal), enabling the activation of the apical amplification mechanism. 

The specification of general network architecture and its parameters can be found in Suppl. Table~\ref{supp_tab:simParams}. 

Additional background inputs, that contribute to setting the network in the awake training, test or sleep states, are provided as distinct stochastic spike trains, with state dependent parameters described respectively by Suppl. Table~\ref{supp_tab:simParamsTraining}, Suppl. Table~\ref{supp_tab:simParamsTest}, and Suppl. Table~\ref{supp_tab:simParamsNREM}.

\subsection{LIF AdEx neuron}

The described network is made of conductance-based leaky-integrate-and-fire (LIF) adaptive exponential (AdEx) neurons, defined by the following coupled equations (\citep{brette2005adaptive}, \citep{gerstner2014neuronal} chapter 6):
\begin{subequations}
    \label{eq:adex}
    \begin{align}
        \begin{split}
            C_{m}\frac{dV}{dt} &= -g_{L}\left( V-E_{L}\right) + g_{L}\,\Delta_{T}\,e^{\left(V-V_{th}\right)/\Delta_{T}} + I_{\mathrm{syn}} - \omega \, ,
        \end{split}
        \label{eq:adex:a}
        \\
        \begin{split}
            \tau_{\omega}\frac{d\omega}{dt} &= a\left(V-E_{L}\right) +b\sum_{k}\delta (t-t_{k}) - \omega \, .
        \end{split}
        \label{eq:adex:b}
    \end{align}
\end{subequations}
Here, the first equation describes the time evolution of the membrane potential~$V$ and incorporates a spike-frequency adaptation (SFA) mechanism through the term $\omega$, whose dynamics is described by the second equation that captures the dependency of the neural excitability on the activation history. Moreover, whenever $V > V_{\mathrm{peak}}$, the membrane potential is set to a reset value $V_{\mathrm{reset}}$.
Here, $C_{m}$ is the membrane capacitance, $g_{L}$ is the membrane leakage conductance, $E_{L}$ the reversal potential, $V_{T}$ the threshold potential, $\Delta_{T}$ the exponential slope parameter in the second right hand term, that models a non linear contribution to the spiking initiation, $\tau_{\omega}$ the adaptation time constant associated with neuronal fatigue, with $a$ and $b$ further adaptation parameters. See Suppl. Table~\ref{supp_tab:simParams} for the simulation parameters.

The input current from excitatory and inhibitory connections, $I_{\mathrm{syn}}$, can be written as:
\begin{equation}
    I_{\mathrm{syn}} = -g_{\mathrm{exc}}(t)\left(V-E_{\mathrm{exc}}\right) + g_{\mathrm{inh}}(t)\left(V-E_{\mathrm{inh}}\right)
\end{equation}
where $g_{\mathrm{exc}}$ and $g_{\mathrm{inh}}$ are the time-dependent excitatory and inhibitory synaptic conductances.
In our model the time evolution of synaptic conductances is shaped according to a peak-normalized $\alpha$ function: 
\begin{equation}
\label{eq:alphaNorm}
    \alpha_{norm}(t-t_s;\tau_s) = \begin{cases} 
    \frac{t-t_s}{\tau_{s}} e^{1-\frac{(t-t_s)}{\tau_{s}}} & \mbox{if } t>t_{s} \\ 0 & \mbox{if } t<t_{s} \end{cases}
\end{equation}
where $\tau_s$ defines the synaptic time scale, and $t_s$ is a presynaptic spike time. An advantage of the $\alpha_{norm}$ formulation is that it yields a peak value equal to $1$ 
for $t-t_s=\tau_s$. The evolution of synaptic conductances is thereby expressed by
\begin{equation}
\label{eq:synEvolution}
    g(t) = W \alpha_{norm}(t - t_s; \tau_s)
\end{equation}
where $W$ is the peak conductance reached at $t-t_s = \tau_s$, that is user-programmable in NEST.

\subsection{Synaptic plasticity}

The evolution of plastic synapses implemented in this model is described by non-linear time-additive Hebbian (NLTAH) spike-timing-dependent plasticity (STDP) \citep{guetig2003learning, morrison2008phenomenological}. The change of each synaptic weight is described by an update function that depends on the times of the spikes of its pre- and post-synaptic neurons $t_{\mathrm{pre}}$ and $t_{\mathrm{post}}$: 

\begin{equation}
    \Delta W = 
    \begin{cases}
        -W_{(-)} \left(\frac{W}{W_{\mathrm{max}}}\right)^{\mu_{(-)}} e^{-(t_{\mathrm{pre}}-t_{\mathrm{post}})/\tau_{(-)}} \, , & \quad \text{if } (t_{\mathrm{pre}}-t_{\mathrm{post}}) > 0\\
        W_{(+)} \left(1 - \frac{W}{W_{\mathrm{max}}}\right)^{\mu_{(+)}} e^{-(t_{\mathrm{pre}}-t_{\mathrm{post}})/\tau_{(+)}} \, , & \quad \text{otherwise}
    \end{cases}
    \label{eq:STDP_plus_minus}
\end{equation}

\noindent

Synaptic weights are upper-bounded by $W_{\mathrm{max}}$ and potentiated or depressed proportionally to $W_{(+)}$ or $W_{(-)}$, respectively. In general, the exponents $\mu_{(+)}$ and $\mu_{(-)}$ can vary in the range $[0,1]$; in the two extreme cases, $\mu_{(\pm)}=0$ and $\mu_{(\pm)}=1$, the model is called additive STDP and multiplicative STDP, respectively. Throughout this work, we use a multiplicative STDP rule, \ie~$\mu_{(+)} = \mu_{(-)} = 1$.

For the purpose of this paper, it is convenient to reformulate Equation \ref{eq:STDP_plus_minus} as below:
\begin{equation}
    \Delta W = 
    \begin{cases}
        -\alpha \, \lambda \, W \, e^{-(t_{\mathrm{pre}}-t_{\mathrm{post}})/\tau_{(-)}} \, , & \quad \text{if } (t_{\mathrm{pre}}-t_{\mathrm{post}}) > 0\\
        \lambda \, (W_{\mathrm{max}} - W) \, e^{-(t_{\mathrm{pre}}-t_{\mathrm{post}})/\tau_{(+)}} \, , & \quad \text{otherwise}
    \end{cases}
    \label{eq:STDP_alfa_lambda}
\end{equation}
where multiplicative parameters have been absorbed by the learning rate $\lambda$ and the synaptic depression/potentiation asymmetry parameter $\alpha$.
According to Hebb's postulate, since the weights of all thalamo-cortical synapses are plastic, if both the input pattern and the contextual signal are kept active for a sufficiently long time, the weights of synapses connecting active thalamic neurons to active cortical neurons will increase. 

\subsection{Training}
\label{sec:methods:training}

During the training phase, all stimuli (perceptual, contextual and background noise) are simultaneously active. The output activity of the thalamic excitatory neurons encodes the image presented as training example. Concurrently, a time-dependent contextual stimulus acts on the cortex, selectively gating the activation of example-specific neuronal assemblies: the contextual signal brings the selected assembly just below the firing threshold -- \ie~ready to be activated by incoming thalamic input -- in accordance with the apical amplification mechanism. Indeed, the perceptual signal propagates from the thalamus to the cortex via plastic feed-forward connections, allowing the cortex to encode memories in the form of neural assemblies shaped by the interaction between contextual input and specific thalamo-cortical synapses. The memory of each training example is sculpted in the connection established among neurons belonging to example-specific cortical assemblies and their thalamo-cortico-thalamic connections. This result is ensured by a hard winner-take-all dynamics (hard-WTA): the substantially higher firing rate, induced in a selected neuronal assembly by the combination of contextual and perceptual signals, drives the inhibitory population to switch off all the other cortical excitatory neurons. 
More specifically, the thalamic layer receives in input a balanced mini-batch of \num{30} examples from the selected dataset (either MNIST or CIFAR-10), corresponding to three examples per class presented sequentially in class order. Each example is presented for \qty{400}{\ms}, during which perceptual input, contextual stimulation, and non-specific noise target thalamic excitatory, cortical excitatory and cortical inhibitory populations respectively. Each presentation is followed by a \qty{400}{\ms} inhibitory-only pause, during which cortical inhibitory neurons are stimulated. The total duration of the training phase sum up to \qty{24}{\s} for each trial. 
A complete specification of the network parameters necessary for simulating this state is provided in Suppl. Table~\ref{supp_tab:simParamsTraining}.

\subsection{Classification}
\label{sec:methods:classification}

During classification, the contextual stimulus is omitted, reducing to zero the prior knowledge injected from outside, while synaptic plasticity is disabled across all connections, ensuring the collection of accuracy metrics in a stationary system. The network operates in an endogenous soft-WTA regime, where neurons previously grouped in multiple example-specific neural assemblies can co-activate, ideally to associate the image presented in classification to the memorized examples belonging to the same class.
Following each sleep epoch, an \qty{8}{\s} pause is applied, during which only the cortical inhibitory population is stimulated. The thalamic layer is subsequently presented with the same balanced mini-batch of 250 examples (25 per class), delivered in class order. Each example is shown for \qty{200}{\ms}, during which only perceptual input to thalamic excitatory population is provided. As in the training phase, each presentation is followed by a \qty{400}{\ms} inhibitory-only pause. The total duration of the classification phase is of 158 s for all the presented examples, out of the pre-classification pause period. A complete specification of the network parameters necessary for simulating this state is provided in Suppl. Table~\ref{supp_tab:simParamsTest}. Details about execution times in Suppl. Table~\ref{supp_tab:execution_times}.

\textbf{Classification accuracy.} \quad During learning, the model encodes perceptual information into partially overlapping thalamo-cortical assemblies through the apical amplification mechanism~\cite{golosio2021thalamocortical}.
Thus, the presentation of new perceptual stimuli promotes the selective activation of training-input-correlated engrams. In this work we estimate the likelihood that the network attributes to each class:
\begin{equation}
    p_{c} \propto \exp{\left(\underset{n \in \{n_c\}}{\max} \; r_n\right)}
\end{equation}
where $\{n_c\}$ is the set of neurons trained to fire when examples from class $c$ are presented, and $\{r_n\}$ are their firing rates during stimulus presentation. The class label estimated as the network's best guess (say $\hat{c}$) is the one that maximizes such probability:
\begin{equation}
    \hat{c} = \argmax_c \; \{p_c\}
\end{equation}
The overall network accuracy $\mathcal{A}$ then comes from the fraction of class labels correctly guessed (\ie,~when $\hat{c}=c$), computed across all test examples:
\begin{equation}
    \mathcal{A} \equiv \braket{\delta_{c,\hat{c}}}_{\mathrm{examples}}
    \label{eq:accuracy}
\end{equation}

\subsection{Sleep}
\label{sec:methods:sleep}

To induce deep-sleep-like oscillatory dynamics, perceptual input is suppressed, and the cortex is driven by a non-specific endogenous Poisson noise, targeting the excitatory population. No contextual signal reaches the cortex, in accordance with the apical isolation principle. In parallel, spike-frequency adaptation of cortical excitatory neurons ($b$ and $\tau_\omega$ parameters in Eq.~(\ref{eq:adex:b})) are increased and cortical inhibitory-to-excitatory synaptic weights are reduced, following a classical modelling recipe for induction of deep-sleep-like oscillation in models \citep{destexhe2009selfsustained}. During the early phase of sleep, orthogonal cortical assemblies alternatively go in up state back-propagating cortical predictions to the thalamus. In turn, the thalamus replays image representation and forwards them to the cortex, thereby promoting the co-activation of cortical assemblies, coherently with a soft-WTA regime, with the set of winners expressed during a single up state correlated to the class of the up state initiator. Through thalamo-cortical and cortico-cortical synaptic plasticity, this unsupervised search process for similar thalamic patterns induces both global homeostatic depression and category-specific associative potentiation across the entire synaptic architecture. These synaptic modifications support a reorganization of thalamo-cortical memories and, ultimately, the emergence of conceptual representations during late-sleep \cite{capone2019sleeplike}. Each sleep epoch consists on 100 s at the end of which all parameters are restored to the default configuration used during training and classification.
A complete specification of the network parameters necessary for simulating this state is provided in Suppl. Table~\ref{supp_tab:simParamsNREM}.

\subsection{Learning, sleep and classification protocols}
\label{sec:methods:protocol}

Data presented in this study, including those shown in Figures~\ref{fig:net_performance} and \ref{fig:net_synapses}, were generated using a predefined learning-classification-sleep protocol applied independently to each trial. The protocol follows the sequence:
\begin{equation}
    \text{training} \rightarrow \text{classification} \rightarrow (100\,\mathrm{s}\,\text{-sleep} \rightarrow \text{classification}) \times \text{20 cycles}
\end{equation}
after the initial training and classification phases, the protocol iteratively repeats, for 20 cycles, a sequence composed of a sleep epoch, a subsequent pause, and a classification phase. This protocol allows network performance to be tracked across successive sleep epochs using the same test set. In total, the full protocol accounts for a duration of \qty{5342}{\s}.
A complete specification of the network parameters used in the different phases is provided in Suppl. Table~\ref{supp_tab:simParams}.

\subsection{Implementation and execution}
\label{sec:methods:implem_exec}

The network has been implemented using NEST 3.8 \cite{gewaltig2007nest, graber2024nest38}, and data analysis and figures are produced using python scripts. 
Simulations run on dual-socket nodes equipped with an AMD EPYC\texttrademark\ \num{7313} CPUs (\num{16} hardware cores per CPU, each SMT-enabled and clocked at \qty{3.0}{\giga\Hz} that can go up to \qty{3.8}{\giga\Hz}). We simultaneously run \num{2} trials per node. The run of a trial takes around \num{9250} wall-clock seconds on the single socket, equivalent to \num{41.1} hardware core-hours (see Suppl. Table~\ref{supp_tab:execution_times}) for the whole learning-sleep protocol described in Sec.~\ref{sec:methods:protocol}. Since the statistics used in this work requires a total of \num{100} trials for each dataset and model plasticity configuration, our estimate for the necessary computing power required to completely reproduce the reported results (MNIST and CIFAR-10, full and partial plasticity) -- on a server setup reasonably equivalent to ours -- amounts to \qty{16.4}{\kilo\corehour}.

In addition, the preprocessing pipeline computed, for each trial and after every sleep epoch, several metrics including: classification accuracy, single-neuron firing rates, single-neuron incoming synaptic weights, cx$\to$cx, cx$\to$th, and th$\to$cx synaptic matrices, as well as example-specific, class-specific, and non-specific categories of synaptic weights. The pipeline also generated low- and high-frequency Gaussian-filtered convolutions of a single entire simulation. Executing the preprocessing script required approximately \num{3290} wall-clock seconds on a single CPU to process \num{100} trials.

Finally, the identification of suitable model parameters required a number of preliminary runs and we estimate we needed about \num{411} additional core-hours for such exploration.

\section*{Declarations}

\subsubsection*{Funding}

Work cofunded by: the European Next Generation EU grants, Italian grants MUR CUP I53C22001400006 (FAIR PE0000013 PNRR) and CUP B51E22000150006 (EBRAINS-Italy IR00011 PNRR); INFN CSN5 grant BRAINSTAIN; APE parallel/distributed lab at INFN Roma. Leonardo Tonielli is a PhD student of the National PhD program in Artificial Intelligence XL cycle, Health and life sciences, organized by Universit\`a Campus Bio-Medico di Roma. Computing resources has been provided by INFN APE LAB (QUonG), CINECA (GALILEO100), and J{\"u}lich Supercomputing Centre (JUWELS).

\subsubsection*{Competing interests}

The authors declare this work has no competing interests.

\subsubsection*{Code availability and reproducibility of results}

The authors support an open science approach in accordance with the FAIR principles (Findability, Accessibility, Interoperability and Reusability).
The full simulation and analysis code supporting the findings of this study, and used to generate all figures, is hosted in a public GitHub repository reachable at \url{https://github.com/APE-group/ThaCo1-FullPlast}.

\subsubsection*{CRediT authorship contribution statement}
\textbf{Leonardo Tonielli:} conceptualization, formal analysis, investigation, methodology, software, visualization, writing - original draft, writing - review \& editing.
\textbf{Cosimo Lupo:} formal analysis, investigation, methodology, software, visualization, writing - original draft, writing - review \& editing, supervision.
\textbf{Elena Pastorelli:} conceptualization, formal analysis, methodology, software, writing - review \& editing, supervision.
\textbf{Giulia De Bonis:} formal analysis, methodology,  writing - review \& editing.
\textbf{Francesco Simula:} software, system administration,  writing - review \& editing.
\textbf{Alessandro Lonardo:} writing - review \& editing, funding acquisition.
\textbf{Pier Stanislao Paolucci:} conceptualization, formal analysis, investigation, methodology, writing - original draft, writing - review \& editing, supervision, funding acquisition.

\bibliography{refs}

\clearpage

\appendix

\setcounter{figure}{0}
\setcounter{table}{0}
\renewcommand{\thefigure}{A.\arabic{figure}}
\renewcommand{\thetable}{A.\arabic{table}}
\renewcommand{\theequation}{A.\arabic{equation}}

\begin{center}
    \LARGE{\textbf{Supplementary Material}}
\end{center}

\begin{figure}[!p]
    \centering
    \includegraphics[width=\textwidth]{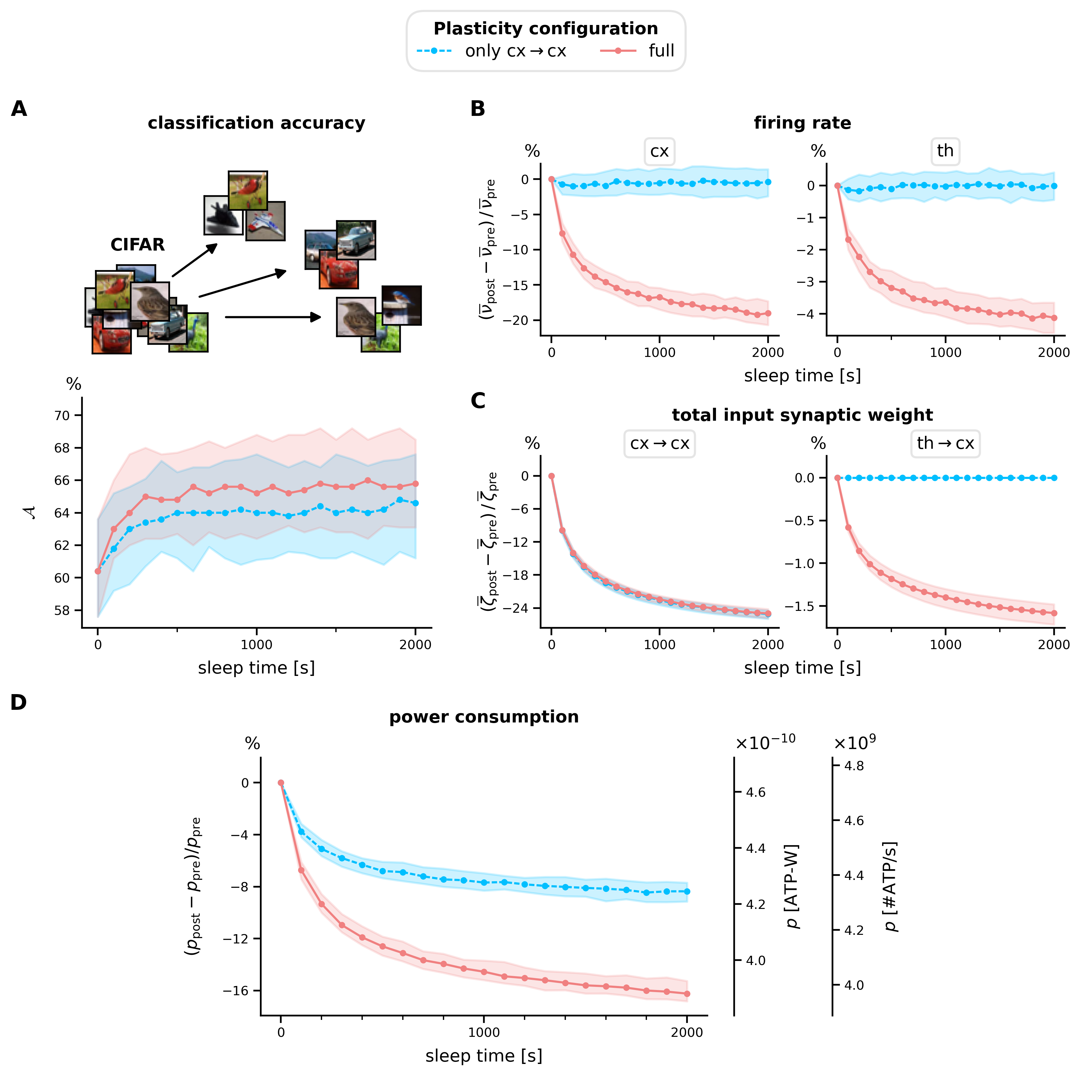}
    \caption{
        \textbf{Sleep effects on cognitive and energetic performance on classification: the CIFAR-10 dataset case.} In analogy to what done in Fig.~\ref{fig:net_performance} for the MNIST dataset, here we focus on CIFAR-10; \textcolor{red}{red} denotes the present full-plasticity model (also th$\to$cx and cx$\to$th plasticity are enabled during sleep), while \textcolor{blue}{blue} refers to the ablated model in which sleep plasticity is restricted to cx$\to$cx synapses only. Observables are measured post-sleep for incremental sleep duration (\num{20} rounds of \qty{100}{\s}-long NREM-sleep). Network is trained on three CIFAR-10 examples per class (\num{10} classes in total). Classification is performed over a balanced test-set composed of \num{250} images.
    }
    \label{supp_fig:net_performance_CIFAR}
\end{figure}

\begin{figure}[!p]
    \centering
    \includegraphics[width=\textwidth]{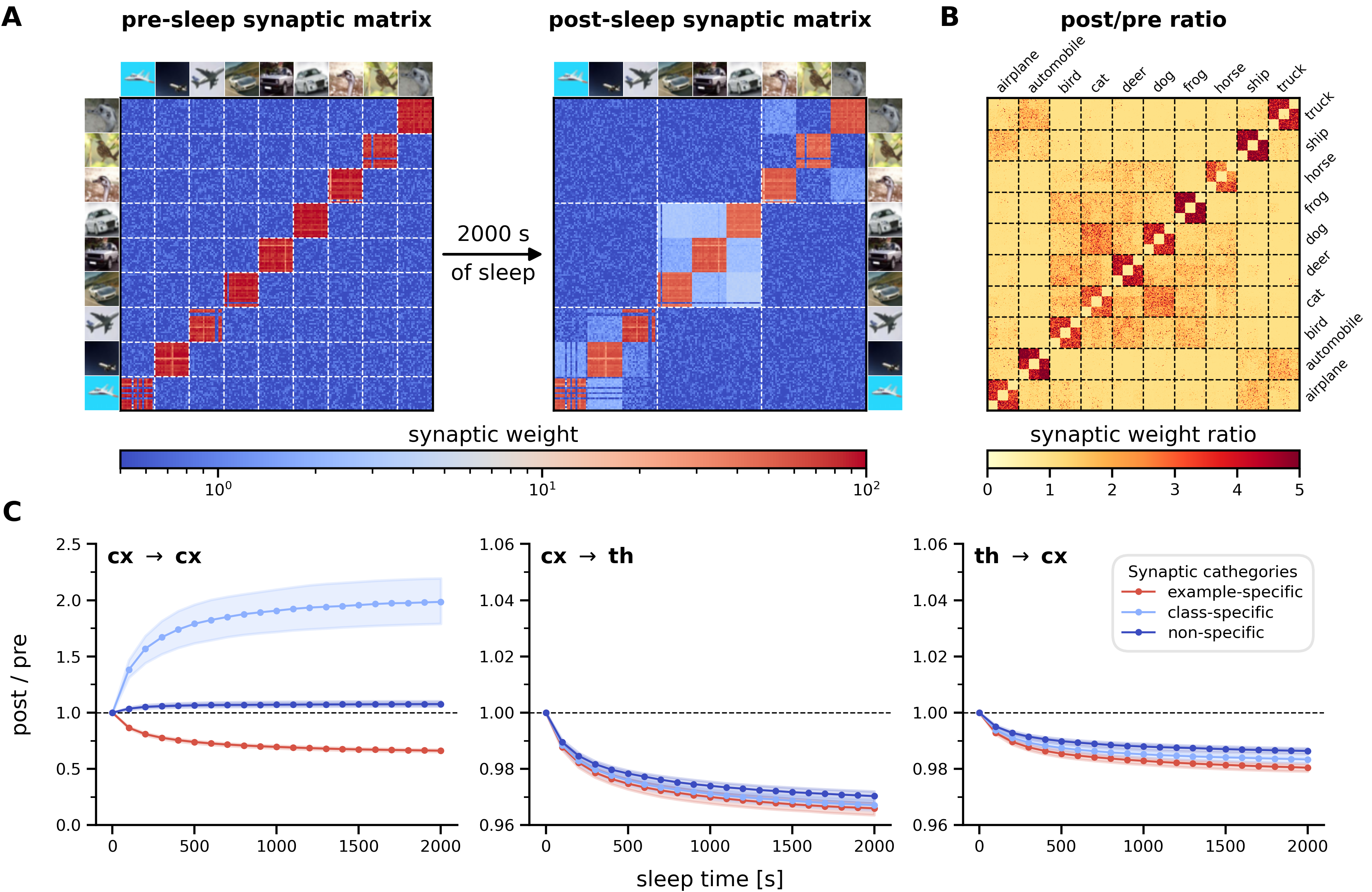}
    \caption{
        \textbf{Network synaptic changes during sleep: the CIFAR-10 dataset case}.
        During sleep, apical isolation mechanism drives an endogenous dynamics inducing associative and homeostatic effects in synaptic weights.
        \textbf{A}) Sleep effects on the cx$\to$cx synaptic matrix for a representative trial, zoomed on the first three image classes, three examples per class. Synaptic weight values are color-coded in log scale (reference weight is \qty{1}{\nano\siemens}). Left: before sleep, stronger synapses (red) emerge within example-specific assemblies, sculpted by apical amplification during the initial training, while cross-example synapses (blue) show no sign of association, staying around the randomly initialized values some orders of magnitude below. Right: after \qty{2000}{\s} of sleep, the strength of example-specific synapses reduces (homeostatic depression; from darker to lighter red), while intermediate-strength synapses spontaneously emerge among cell assemblies encoding for the same digit class (memory association; light blue).
        \textbf{B}) Trial-average (\num{100} independent trials) of the entire post-sleep/pre-sleep cx$\to$cx synaptic matrix (ten image classes), coherently showing the homeostatic and associative effects previously reported for a single trial. In addition, a multi-level hierarchy structure emerges, correlating entries from different -- but homogeneous -- classes (\eg, all animals together, and all vehicles together), and even more finely inside such macro-categories (\eg,~cats and dogs, automobiles and trucks), coherently with the similarity structure encoded in the preprocessed dataset, Fig.~\ref{fig:preprocessing}D.
        \textbf{C)} Homeostatic effects for example-specific synapses and associative effects for class-specific synapses shown as a function of sleep time for for the connectivity matrices: cx$\to$cx, cx$\to$th and th$\to$cx; data are computed as the ratio between post-sleep/pre-sleep averages over the three functional classes of connectivity for every given trial; medians and inter-quartile ranges over the means from the \num{100} independent trials are shown.
    }
    \label{supp_fig:net_synapses_CIFAR}
\end{figure}

\noindent
This section contains a set of supplementary figures, focusing on: 1) the accuracy and energetic improvements of CIFAR-10 input due to deep-sleep-like activity in ThaCo, Fig.~\ref{supp_fig:net_performance_CIFAR}; 2) the synaptic reorganization induced by sleep on CIFAR-10, Fig.~\ref{supp_fig:net_synapses_CIFAR}.

The values of all parameters necessary for reproducing the simulations and retrieve the results presented in this paper are reported in a series of tables. In particular, in Table~\ref{supp_tab:simParams} are reported general settings, while brain-state-specific parameters can be found in Table~\ref{supp_tab:simParamsTraining} for awake training, in Table~\ref{supp_tab:simParamsTest} for awake classification, and in Table~\ref{supp_tab:simParamsNREM} for NREM-like sleep. Finally, Table~\ref{supp_tab:synapses} contains numerical values for the initialization of synapses and for their evolution through the training-sleep protocol.

\begingroup
\setlength{\tabcolsep}{6pt}
\renewcommand{\arraystretch}{1.25}
\begin{table}[p]
    \centering
    \caption{
        \textbf{Input datasets, training times and neural population sizes.} See Sec.~\ref{sec:methods:input} for an explanation of the techniques used to preprocess examples from the MNIST and CIFAR-10 datasets and administer them to the network. See Sec.~\ref{sec:methods:architecture} for the explanation of the multi-layer plastic spiking network architecture used in this work.
        Presentation and pause times correspond to the input stimulus intervals used during training and classification, respectively, whereas the relaxation time corresponds to the inter-stage pre-classification pause.
    }
    \label{supp_tab:simParams}

    \bigskip

    \begin{tabularx}{\textwidth}{C{0.30}L{2.00}C{1.40}C{0.30}}
        \multicolumn{4}{c}{\textbf{Input}} \\
        \midrule
        \rowcolor{gray!15}
        &
            Dataset type &
            MNIST or CIFAR-10 &
            \\
        &
            Number of classes &
            \num{10} &
            \\
        \rowcolor{gray!15}
        &
            Number of encoded features &
            81 &
            \\
        &
            Length of input vector after hot-encoding &
            \num{324} &
            \\
        \rowcolor{gray!15}
        &
            Number of training examples per class &
            \num{3} &
            \\
        &
            Number of classification examples per class &
            \num{25} &
            \\
        \rowcolor{gray!15}
        &
            Training presentation time &
            \qty{400}{\ms} &
            \\
        &
            Classification presentation time &
            \qty{200}{\ms} &
            \\
        \rowcolor{gray!15}
        &
            Training or classification pause time &
            \qty{400}{\ms} &
            \\
        &
            Pre-classification relaxation time &
            \qty{8}{\s} &
            \\
    \end{tabularx}

    \bigskip

    \begin{tabularx}{\textwidth}{C{0.30}L{2.00}C{1.40}C{0.30}}
        \multicolumn{4}{c}{\textbf{Network population sizes}}     \\
        \midrule
        \rowcolor{gray!15}
        &
            Number of cx exc neurons &
            \num{600} &
            \\
        &
            Number of cx inh neurons &
            \num{200} &
            \\
        \rowcolor{gray!15}
        &
            Number of cx exc neurons per assembly &
            \num{20} &
            \\
        &
            Number of th exc neurons & 
            \num{324} &
            \\
        \rowcolor{gray!15}
        &
            Number of th inh neurons &
            \num{200} &
            \\
    \end{tabularx}
\end{table}
\endgroup

\begingroup
\setlength{\tabcolsep}{6pt}
\renewcommand{\arraystretch}{1.25}
\begin{table}[p]
    \centering
    \caption{
        \textbf{Awake Training.}
        Neuron parameters refer to ``aeif\_cond\_alpha'' neuron model in NEST; parameters not explicitly specified here, correspond to the default values of NEST, version 3.8.
        The explanation of the encoding of external stimuli and pauses between them by means of Poissonian noise can be found in Sec.~\ref{sec:methods:training}.
        Accordingly, example rate and pause rate columns indicate the stimulus rates used for the image presentation in training, whereas the relaxation rate refers to the inter-stage pre-training pause.
        Synaptic plasticity is implemented via ``stdp\_synapse'' NEST model.
    }
    \label{supp_tab:simParamsTraining}

    \bigskip
    
    \begin{tabularx}{\textwidth}{C{0.35}L{1.1}C{1.5}C{0.75}L{1.1}C{1.5}C{0.75}L{1.1}C{1.5}C{0.35}}
        \rowcolor{gray!0}
        \multicolumn{10}{c}{\textbf{Neuron parameters}} \\
        \midrule
        \rowcolor{gray!15}
        &
            a &
            \qty{4}{\nano\siemens} &
            &
            b &
            \qty{0.01}{\pico\ampere} &
            &
            $C_{\mathrm{m}}$ &
            \qty{281}{\pico\farad} &
            \\
        \rowcolor{gray!0}
        &
            $\Delta_{\mathrm{T}}$ &
            \qty{2}{\milli\volt} &
            &
            $E_{\mathrm{L}}$ &
            \qty{-70.6}{\milli\volt} &
            &
            $E_{\mathrm{exc}}$ &
            \qty{0}{\milli\volt} &
            \\
        \rowcolor{gray!15}
        &
            $E_{\mathrm{inh}}$ &
            \qty{-85}{\milli\volt} &
            &
            $g_{\mathrm{L}}$ &
            \qty{30}{\nano\siemens} &
            &
            $I_{\mathrm{e}}$ &
            \qty{0}{\pico\ampere} &
            \\
        \rowcolor{gray!0}
        &
            $t_{\mathrm{ref}}$ &
            \qty{2}{\milli\s} &
            &
            $\tau_{\mathrm{syn,exc}}$ &
            \qty{0.2}{\milli\s} &
            &
            $\tau_{\mathrm{syn,inh}}$ &
            \qty{2}{\milli\s} &
            \\
        \rowcolor{gray!15}
        &
            $\tau_{\mathrm{w}}$ &
            \qty{144}{\milli\s} &
            &
            $V_{\mathrm{m}}$ &
            \qty{-71.2}{\milli\volt} &
            &
            $V_{\mathrm{peak}}$ &
            \qty{0}{\milli\volt} &
            \\
        \rowcolor{gray!0}
        &
            $V_{\mathrm{reset}}$ &
            \qty{-60}{\milli\volt} &
            &
            $V_{\mathrm{th}}$ &
            \qty{-50.4}{\milli\volt} &
            &
            &
            &
            \\
    \end{tabularx}

    \bigskip

    \begin{tabularx}{\textwidth}{L{0.7}C{1.1}C{1.1}C{1.1}C{1}C{1}}
        \multicolumn{6}{c}{\textbf{Noise parameters}} \\
        \midrule
        &
            example rate &
            pause rate &
            relaxation rate &
            W &
            delay \\
        \midrule
        \rowcolor{gray!15}
        th exc &
            \qty{30}{\kilo\Hz} &
            0 &
            0 &
            \qty{8}{\nano\siemens} &
            \qty{1}{\milli\s} \\
        th inh &
            -- &
            -- &
            -- &
            -- &
            -- \\
        \rowcolor{gray!15}
        cx exc &
            \qty{2}{\kilo\Hz} &
            0 &
            0 &
            \qty{15}{\nano\siemens} &
            \qty{3}{\milli\s} \\
        cx inh &
            \qty{10}{\kilo\Hz} &
            \qty{40}{\kilo\Hz} &
            0 &
            \qty{3}{\nano\siemens} &
            \qty{1}{\milli\s} \\
    \end{tabularx}
    
    \bigskip

    \begin{tabularx}{\textwidth}{XC{1}C{1}C{1}C{1}C{1}C{1}}
        \multicolumn{7}{c}{\textbf{Synaptic parameters}} \\
        \midrule
        &
            $W_{0}$ &
            $W_{\mathrm{max}}$ &
            delay &
            $\lambda$ &
            $\alpha$ &
            $\mu_{(\pm)}$ \\
        \midrule
        \rowcolor{gray!15}
        th$\to$cx &
            \qty{1}{\nano\siemens} &
            \qty{5.5}{\nano\siemens} &
            \qty{1}{\milli\s} &
            \num{0.03} &
            \num{1} &
            \num{1} \\
        th exc-exc &
            -- &
            -- &
            -- &
            -- &
            -- &
            -- \\
        \rowcolor{gray!15}
        th exc-inh &
            \qty{10}{\nano\siemens} &
            -- &
            \qty{1}{\milli\s} &
            -- &
            -- &
            -- \\
        th inh-exc &
            \qty{-1}{\nano\siemens} &
            -- &
            \qty{1}{\milli\s} &
            -- &
            -- &
            -- \\
        \rowcolor{gray!15}
        th inh-inh &
            -- &
            -- &
            -- &
            -- &
            -- &
            -- \\
        cx$\to$th &
            \qty{1}{\nano\siemens} &
            \qty{130}{\nano\siemens} &
            \qty{1}{\milli\s} &
            \num{0.08} &
            \num{1} &
            \num{1} \\
        \rowcolor{gray!15}
        cx exc-exc &
            \qty{0.5}{\nano\siemens} &
            \qty{150}{\nano\siemens} &
            \qty{1}{\milli\s} &
            0.12 &
            1 &
            1 \\
        cx exc-inh &
            \qty{60}{\nano\siemens} &
            -- &
            \qty{1}{\milli\s} &
            -- &
            -- &
            -- \\
        \rowcolor{gray!15}
        cx inh-exc &
            \qty{-4}{\nano\siemens} &
            -- &
            \qty{1}{\milli\s} &
            -- &
            -- &
            -- \\
        cx inh-inh &
            \qty{-4}{\nano\siemens} &
            -- &
            \qty{1}{\milli\s} &
            -- &
            -- &
            -- \\
    \end{tabularx}
    
\end{table}
\endgroup


\begingroup
\setlength{\tabcolsep}{6pt}
\renewcommand{\arraystretch}{1.25}
\begin{table}[p]
    \centering
    \caption{
        \textbf{Awake Classification.}
        Neuron parameters refer to ``aeif\_cond\_alpha'' neuron model in NEST; parameters not explicitly specified here, correspond to the default values of NEST, version 3.8.
        The explanation of the encoding of external stimuli and pauses between them by means of Poissonian noise can be found in Sec.~\ref{sec:methods:classification}.
        Accordingly, example rate and pause rate columns indicate the stimulus rates used for the image presentation in classification, whereas the relaxation rate refers to the inter-stage pre-classification pause.
        Synaptic plasticity is implemented via ``stdp\_synapse'' NEST model.
    }
    \label{supp_tab:simParamsTest}

    \bigskip

    \begin{tabularx}{\textwidth}{C{0.35}L{1.1}C{1.5}C{0.75}L{1.1}C{1.5}C{0.75}L{1.1}C{1.5}C{0.35}}
        \rowcolor{gray!0}
        \multicolumn{10}{c}{\textbf{Neuron parameters}} \\
        \midrule
        \rowcolor{gray!15}
        &
            a &
            \qty{4}{\nano\siemens} &
            &
            b &
            \qty{0.01}{\pico\ampere} &
            &
            $C_{\mathrm{m}}$ &
            \qty{281}{\pico\farad} &
            \\
        \rowcolor{gray!0}
        &
            $\Delta_{\mathrm{T}}$ &
            \qty{2}{\milli\volt} &
            &
            $E_{\mathrm{L}}$ &
            \qty{-70.6}{\milli\volt} &
            &
            $E_{\mathrm{exc}}$ &
            \qty{0}{\milli\volt} &
            \\
        \rowcolor{gray!15}
        &
            $E_{\mathrm{inh}}$ &
            \qty{-85}{\milli\volt} &
            &
            $g_{\mathrm{L}}$ &
            \qty{30}{\nano\siemens} &
            &
            $I_{\mathrm{e}}$ &
            \qty{0}{\pico\ampere} &
            \\
        \rowcolor{gray!0}
        &
            $t_{\mathrm{ref}}$ &
            \qty{2}{\milli\s} &
            &
            $\tau_{\mathrm{syn,exc}}$ &
            \qty{0.2}{\milli\s} &
            &
            $\tau_{\mathrm{syn,inh}}$ &
            \qty{2}{\milli\s} &
            \\
        \rowcolor{gray!15}
        &
            $\tau_{\mathrm{w}}$ &
            \qty{144}{\milli\s} &
            &
            $V_{\mathrm{m}}$ &
            \qty{-71.2}{\milli\volt} &
            &
            $V_{\mathrm{peak}}$ &
            \qty{0}{\milli\volt} &
            \\
        \rowcolor{gray!0}
        &
            $V_{\mathrm{reset}}$ &
            \qty{-60}{\milli\volt} &
            &
            $V_{\mathrm{th}}$ &
            \qty{-50.4}{\milli\volt} &
            &
            &
            &
            \\
    \end{tabularx}
    
    \bigskip

    \begin{tabularx}{\textwidth}{L{0.7}C{1.1}C{1.1}C{1.1}C{1}C{1}}
        \multicolumn{6}{c}{\textbf{Noise parameters}} \\
        \midrule
        &
            example rate &
            pause rate &
            relaxation rate &
            W &
            delay \\
        \midrule
        \rowcolor{gray!15}
        th exc &
            \qty{30}{\kilo\Hz} &
            0 &
            0 &
            \qty{8}{\nano\siemens} &
            \qty{1}{\milli\s} \\
        th inh &
            -- &
            -- &
            -- &
            -- &
            -- \\
        \rowcolor{gray!15}
        cx exc &
            0 &
            0 &
            0 &
            \qty{15}{\nano\siemens} &
            \qty{3}{\milli\s} \\
        cx inh &
            0 &
            \qty{20}{\kilo\Hz} &
            \qty{40}{\kilo\Hz} &
            \qty{3}{\nano\siemens} &
            \qty{1}{\milli\s} \\
    \end{tabularx}
    
    \bigskip

    \begin{tabularx}{\textwidth}{XC{1}C{1}C{1}C{1}C{1}C{1}}
        \multicolumn{7}{c}{\textbf{Synaptic parameters}} \\
        \midrule
        &
            $W_{0}$ &
            $W_{\mathrm{max}}$ &
            delay &
            $\lambda$ &
            $\alpha$ &
            $\mu_{(\pm)}$ \\
        \midrule
        \rowcolor{gray!15}
        th$\to$cx &
            \qty{1}{\nano\siemens} &
            \qty{5.5}{\nano\siemens} &
            \qty{1}{\milli\s} &
            \num{0} &
            \num{1} &
            \num{1} \\
        th exc-exc &
            -- &
            -- &
            -- &
            -- &
            -- &
            -- \\
        \rowcolor{gray!15}
        th exc-inh &
            \qty{10}{\nano\siemens} &
            -- &
            \qty{1}{\milli\s} &
            -- &
            -- &
            -- \\
        th inh-exc &
            \qty{-1}{\nano\siemens} &
            -- &
            \qty{1}{\milli\s} &
            -- &
            -- &
            -- \\
        \rowcolor{gray!15}
        th inh-inh &
            -- &
            -- &
            -- &
            -- &
            -- &
            -- \\
        cx$\to$th &
            \qty{1}{\nano\siemens} &
            \qty{130}{\nano\siemens} &
            \qty{1}{\milli\s} &
            \num{0} &
            \num{1} &
            \num{1} \\
        \rowcolor{gray!15}
        cx exc-exc &
            \qty{0.5}{\nano\siemens} &
            \qty{150}{\nano\siemens} &
            \qty{1}{\milli\s} &
            0 &
            1 &
            1 \\
        cx exc-inh &
            \qty{60}{\nano\siemens} &
            -- &
            \qty{1}{\milli\s} &
            -- &
            -- &
            -- \\
        \rowcolor{gray!15}
        cx inh-exc &
            \qty{-4}{\nano\siemens} &
            -- &
            \qty{1}{\milli\s} &
            -- &
            -- &
            -- \\
        cx inh-inh &
            \qty{-4}{\nano\siemens} &
            -- &
            \qty{1}{\milli\s} &
            -- &
            -- &
            -- \\
    \end{tabularx}
    
\end{table}
\endgroup


\begingroup
\setlength{\tabcolsep}{6pt}
\renewcommand{\arraystretch}{1.25}
\begin{table}[p]
    \centering
    \caption{
        \textbf{NREM.}
        Sleep epochs denote the checkpointing procedure used to benchmark awake classification performance across successive sleep periods. The thermalization time corresponds to a preparatory interval preceding NREM sleep, during which STDP plasticity is disabled.
        Neuron parameters refer to ``aeif\_cond\_alpha'' neuron model in NEST; parameters not explicitly specified here, correspond to the default values of NEST, version 3.8.
        The explanation of which parameters are crucial to set the network in an NREM-like SO regime can be found in Sec.~\ref{sec:methods:sleep}. Here, the complete set of parameters is included, comprising those not changed from awake states.
        Synaptic plasticity is implemented via ``stdp\_synapse'' NEST model.
    }
    \label{supp_tab:simParamsNREM}

    \bigskip

    \begin{tabularx}{\textwidth}{C{0.60}L{1.50}C{1.30}C{0.60}}
        \multicolumn{4}{c}{\textbf{Sleep parameters}} \\
        \midrule
        \rowcolor{gray!15}
        &
            Number of sleep epochs &
            \num{20} &
            \\
        &
            Thermalization time per epoch &
            \qty{10}{\s} &
            \\
        \rowcolor{gray!15}
        &
            Sleep time per cycle &
            \qty{100}{\s} &
            \\
    \end{tabularx}

    \bigskip

    \begin{tabularx}{\textwidth}{C{0.35}L{1.1}C{1.5}C{0.75}L{1.1}C{1.5}C{0.75}L{1.1}C{1.5}C{0.35}}
        \rowcolor{gray!0}
        \multicolumn{10}{c}{\textbf{Neuron parameters}} \\
        \midrule
        \rowcolor{gray!15}
        &
            a &
            \qty{4}{\nano\siemens} &
            &
            b &
            \qty{120}{\pico\ampere} &
            &
            $C_{\mathrm{m}}$ &
            \qty{281}{\pico\farad} &
            \\
        \rowcolor{gray!0}
        &
            $\Delta_{\mathrm{T}}$ &
            \qty{2}{\milli\volt} &
            &
            $E_{\mathrm{L}}$ &
            \qty{-70.6}{\milli\volt} &
            &
            $E_{\mathrm{exc}}$ &
            \qty{0}{\milli\volt} &
            \\
        \rowcolor{gray!15}
        &
            $E_{\mathrm{inh}}$ &
            \qty{-85}{\milli\volt} &
            &
            $g_{\mathrm{L}}$ &
            \qty{30}{\nano\siemens} &
            &
            $I_{\mathrm{e}}$ &
            \qty{0}{\pico\ampere} &
            \\
        \rowcolor{gray!0}
        &
            $t_{\mathrm{ref}}$ &
            \qty{2}{\milli\s} &
            &
            $\tau_{\mathrm{syn,exc}}$ &
            \qty{0.2}{\milli\s} &
            &
            $\tau_{\mathrm{syn,inh}}$ &
            \qty{2}{\milli\s} &
            \\
        \rowcolor{gray!15}
        &
            $\tau_{\mathrm{w}}$ &
            \qty{400}{\milli\s} &
            &
            $V_{\mathrm{m}}$ &
            \qty{-71.2}{\milli\volt} &
            &
            $V_{\mathrm{peak}}$ &
            \qty{0}{\milli\volt} &
            \\
        \rowcolor{gray!0}
        &
            $V_{\mathrm{reset}}$ &
            \qty{-60}{\milli\volt} &
            &
            $V_{\mathrm{th}}$ &
            \qty{-50.4}{\milli\volt} &
            &
            &
            &
            \\
    \end{tabularx}

    \bigskip

    \begin{tabularx}{\textwidth}{C{0.75}L{1.2}C{1.1}C{1.1}C{1.1}C{0.75}}
        \multicolumn{6}{c}{\textbf{Noise parameters}} \\
        \midrule
        &
            &
            rate &
            W &
            delay &
            \\
        \midrule
        \rowcolor{gray!15}
        &
            th exc &
            0 &
            \qty{8}{\nano\siemens} &
            \qty{1}{\milli\s} &
            \\
        &
            th inh &
            -- &
            -- &
            -- &
            \\
        \rowcolor{gray!15}
        &
            cx exc &
            \qty{710}{\kilo\Hz} &
            \qty{15}{\nano\siemens} &
            \qty{3}{\milli\s} &
            \\
        &
            cx inh &
            0 &
            \qty{3}{\nano\siemens} &
            \qty{1}{\milli\s} &
            \\
    \end{tabularx}
    
    \bigskip

    \begin{tabularx}{\textwidth}{XC{1}C{1}C{1}C{1}C{1}C{1}}
        \multicolumn{7}{c}{\textbf{Synaptic parameters}} \\
        \midrule
        &
            $W_{0}$ &
            $W_{\mathrm{max}}$ &
            delay &
            $\lambda$ &
            $\alpha$ &
            $\mu_{(\pm)}$ \\
        \midrule
        \rowcolor{gray!15}
        th$\to$cx &
            \qty{1}{\nano\siemens} &
            \qty{5.5}{\nano\siemens} &
            \qty{1}{\milli\s} &
            \num{2e-7} &
            \num{20} &
            \num{1} \\
        th exc-exc &
            -- &
            -- &
            -- &
            -- &
            -- &
            -- \\
        \rowcolor{gray!15}
        th exc-inh &
            \qty{10}{\nano\siemens} &
            -- &
            \qty{1}{\milli\s} &
            -- &
            -- &
            -- \\
        th inh-exc &
            \qty{-1}{\nano\siemens} &
            -- &
            \qty{1}{\milli\s} &
            -- &
            -- &
            -- \\
        \rowcolor{gray!15}
        th inh-inh &
            -- &
            -- &
            -- &
            -- &
            -- &
            -- \\
        cx$\to$th &
            \qty{1}{\nano\siemens} &
            \qty{130}{\nano\siemens} &
            \qty{1}{\milli\s} &
            \num{2e-7} &
            \num{20} &
            \num{1} \\
        \rowcolor{gray!15}
        cx exc-exc &
            \qty{0.5}{\nano\siemens} &
            \qty{150}{\nano\siemens} &
            \qty{1}{\milli\s} &
            \num{2e-6} &
            \num{20} &
            \num{1} \\
        cx exc-inh &
            \qty{60}{\nano\siemens} &
            -- &
            \qty{1}{\milli\s} &
            -- &
            -- &
            -- \\
        \rowcolor{gray!15}
        cx inh-exc &
            \qty{-0.5}{\nano\siemens} &
            -- &
            \qty{1}{\milli\s} &
            -- &
            -- &
            -- \\
        cx inh-inh &
            \qty{-4}{\nano\siemens} &
            -- &
            \qty{1}{\milli\s} &
            -- &
            -- &
            -- \\
    \end{tabularx}
    
\end{table}
\endgroup

\begingroup
\setlength{\tabcolsep}{6pt}
\renewcommand{\arraystretch}{1.25}
\begin{table}[p]
    \centering
    \caption{
        \textbf{Synaptic changes from pre- to post-sleep, for MNIST dataset.} Pre-sleep synaptic efficacy and post- to pre-sleep ratio for cx$\to$cx, cx$\to$th and th$\to$cx connections (mean and SEM over \num{100} independent trials). Pre-training weights express the initialization values for synapses: cx$\to$cx are drawn from a uniform distribution, while cx$\to$th and th$\to$cx are initialized to \qty{1}{\nano\siemens} and, thanks to the apical amplification selective mechanism, remain unchanged after the training notwithstanding their plasticity for class-specific and non-specific connections; then, they are affected by the sleep-induced synaptic reorganization.
    }
    \label{supp_tab:synapses}

    \bigskip

    \begin{tabularx}{\textwidth}{L{1}C{1}C{1}C{1}}
        \multicolumn{4}{c}{\textbf{Pre-training weights} $\boldsymbol{[\unit{\nano\siemens}]}$} \\
        &
            example-specific &
            class-specific &
            non-specific \\
        \midrule
        \rowcolor{gray!15}
        cx$\to$cx synapses &
            $\mathrm{Unif([0.001, 1])}$ &
            $\mathrm{Unif([0.001, 1])}$ &
            $\mathrm{Unif([0.001, 1])}$ \\
        cx$\to$th synapses &
            \num{1} &
            \num{1} &
            \num{1} \\
        \rowcolor{gray!15}
        th$\to$cx synapses &
            \num{1} &
            \num{1} &
            \num{1} \\
    \end{tabularx}

    \bigskip

    \begin{tabularx}{\textwidth}{L{1}C{1}C{1}C{1}}
        \multicolumn{4}{c}{\textbf{Pre-sleep weights} $\boldsymbol{[\unit{\nano\siemens}]}$} \\
        &
            example-specific &
            class-specific &
            non-specific \\
        \midrule
        \rowcolor{gray!15}
        cx$\to$cx synapses &
            \num{64.6(2)} &
            \num{0.5004(2)} & 
            \num{0.5005(1)} \\
        cx$\to$th synapses &
            \num{47.0(1)} &
            \num{1} &
            \num{1} \\
        \rowcolor{gray!15}
        th$\to$cx synapses &
            \num{2.600(3)} &
            \num{1} &
            \num{1} \\
    \end{tabularx}

    \bigskip

    \begin{tabularx}{\textwidth}{L{1}C{1}C{1}C{1}}
        \multicolumn{4}{c}{\textbf{Post- to pre-sleep weights ratio}} \\
        &
            example-specific &
            class-specific &
            non-specific \\
        \midrule
        \rowcolor{gray!15}
        cx$\to$cx synapses &
            \num{0.673(2)} &
            \num{1.802(3)} &
            \num{1.075(5)} \\
        cx$\to$th synapses &
            \num{0.9688(3)} &
            \num{1.0133(6)} &
            \num{1.0026(2)} \\
        \rowcolor{gray!15}
        th$\to$cx synapses &
            \num{0.9697(3)} &
            \num{0.99807(8)} &
            \num{0.99962(3)} \\
        \end{tabularx}
\end{table}
\endgroup

\begingroup
    \setlength{\tabcolsep}{6pt}
    \renewcommand{\arraystretch}{1.25}
    \begin{table}[p]
        \centering
        \caption{
            \textbf{Simulation times.}
            Biological times indicate the simulated duration of each stage or epoch (training, classification, and sleep). Wall-clock times refer to the corresponding computational times required to simulate the system and to save data required to generate figures presented in this study.
            See Sec.~\ref{sec:methods:implem_exec} for a description of the software and hardware execution environment. See Sec.~\ref{sec:methods:training} for the explanation of the training phase, Sec.~\ref{sec:methods:classification} for the classification phase, and Sec.~\ref{sec:methods:sleep} for the sleep phase. See Sec.~\ref{sec:methods:protocol} for details about the alternation between epochs of sleep and post-sleep classification stages and about the trials to accumulate statistics.
        }
        \label{supp_tab:execution_times}
        
        \begin{tabularx}{\textwidth}{C{0.2}L{2.4}C{1.2}C{0.2}}
            \multicolumn{4}{c}{\textbf{Simulation times per trial}} \\
            \midrule
            &
                Number of simultaneous trial executions per node &
                \num{2} &
                \\
            \rowcolor{gray!15}
            &
                Number of hardware cores per trial &
                \num{16} &
                \\
            &
                Number of SMT threads per core &
                \num{2} &
                \\
            \rowcolor{gray!15}
            &
                Training time (biological) per trial &
                \qty{24}{\s} &
                \\
            &
                Training time (wall-clock) per trial &
                \qty{50}{\s} &
                \\
            \rowcolor{gray!15}
            &
                Classification time (biological) per sleep epoch &
                \qty{158}{\s} &
                \\
            &
                Classification time (wall-clock) per sleep epoch &
                \qty{200}{\s} &
                \\
            \rowcolor{gray!15}
            &
                Sleep time (biological) per sleep epoch per trial &
                \qty{100}{\s} &
                \\
            &
                Sleep time (wall-clock) per sleep epoch per trial &
                \qty{260}{\s} &
                \\
            \rowcolor{gray!15}
            &
                Classification epochs per trial &
                \num{21} &
                \\
            &
                Sleep epochs per trial &
                \num{20} &
                \\
            \rowcolor{gray!15}
            &
                Total time (biological) per trial &
                \qty{5342}{\s} &
                \\
            &
                Total time (wall-clock) per trial &
                \qty{9250}{\s} &
                \\
            \rowcolor{gray!15}
            &
                Total (hardware) corehours per trial &
                \qty{41.1}{\corehour} &
                \\
        \end{tabularx}

        \bigskip
        
        \begin{tabularx}{\textwidth}{C{0.2}L{2.4}C{1.2}C{0.2}}
            \multicolumn{4}{c}{\textbf{Total simulation time}} \\
            \midrule
            &
                Number of datasets &
                \num{2} &
                \\
            \rowcolor{gray!15}
            &
                Number of plasticity configurations per dataset &
                \num{2} &
                \\
            &
                Number of trials per dataset and configuration &
                \num{100} &
                \\
            \rowcolor{gray!15}
            &
                Total (hardware) corehours &
                \qty{16.4}{\kilo\corehour} &
                \\
        \end{tabularx}
    \end{table}
\endgroup

\end{document}